\documentclass[aps,prb,showpacs,eqsecnum,twocolumn,superscriptaddress,amsmath,amssymb]{revtex4-1}

\usepackage[utf8]{inputenc}

\usepackage{graphicx}
\usepackage[colorlinks=true,citecolor=blue]{hyperref}
\usepackage{hypernat}
\usepackage{color}%only needed for coloured text
 
\usepackage{verbatim} %mehrzeilige Kommentare

\newcommand{\bra}[1]{\langle #1|}
\newcommand{\ket}[1]{|#1\rangle}
\newcommand{\RPG}[1]{\re\Psi\left(\frac{1}{2}+i\frac{\beta#1}{2\pi}\right)} %real part of polygamma function of 1/2+i\beta/(2pi)*x
\renewcommand{\vec}[1]{\mathbf{#1}}
\DeclareMathOperator{\re}{\text{Re}}

\def\up{\uparrow}
\def\down{\downarrow}
\def\kB {k_\text{B}}

\begin{document}

\title{Influence of spin waves on transport through a quantum-dot spin valve}

\author{Bj\"orn Sothmann}
\affiliation{Theoretische Physik, Universit\"at Duisburg-Essen and CeNIDE, 47048 Duisburg, Germany}
\author{J\"urgen K\"onig}
\affiliation{Theoretische Physik, Universit\"at Duisburg-Essen and CeNIDE, 47048 Duisburg, Germany}
\author{Anatoli Kadigrobov}
\affiliation{Departement of Physics, G\"oteborg University, 412 96 G\"oteborg, Sweden}
\affiliation{Theoretische Physik III, Ruhr-Universit\"at Bochum, 44780 Bochum, Germany}

\date{\today}

\begin{abstract}
We study the influence of spin waves on transport through a single-level quantum dot weakly coupled to ferromagnetic electrodes with noncollinear magnetizations. Side peaks appear in the differential conductance due to emission and absorption of spin waves. We, furthermore, investigate the nonequilibrium magnon distributions generated in the source and drain lead. In addition, we show how magnon-assisted tunneling can generate a fully spin-polarized current without an applied transport voltage. We discuss the influence of spin waves on the current noise. Finally, we show how the magnonic contributions to the exchange field can be detected in the finite-frequency Fano factor.
\end{abstract}

%73.23.Hk 	Coulomb blockade; single-electron tunneling
%72.25.Mk 	Spin transport through interfaces
%85.75.-d 	Magnetoelectronics; spintronics: devices exploiting spin polarized transport or integrated magnetic fields
%72.70.+m 	Noise processes and phenomena
%73.63.Kv 	Quantum dots

\maketitle
\section{\label{sec:intro}Introduction}
Quantum dots coupled to ferromagnetic electrodes have generated much interest in the recent past due to their possible applications in spintronics. 
Experimentally realizations of such systems include small metallic grains,~\cite{deshmukh_using_2002,bernand-mantel_evidence_2006,bernand-mantel_anisotropic_2009} semiconductor quantum dots,~\cite{hamaya_spin_2007,hamaya_electric-field_2007,hamaya_kondo_2007,hamaya_oscillatory_2008,hamaya_tunneling_2008,hamaya_spin-related_2009} single-wall carbon nanotubes,~\cite{jensen_single-wall_2003,sahoo_electric_2005,liu_spin-dependent_2006,hauptmann_electric-field-controlled_2008,merchant_current_2009} as well as single molecules~\cite{pasupathy_kondo_2004} contacted by ferromagnetic electrodes.
From a theoretical point of view, quantum-dot spin valves, i.e., quantum dots coupled to noncollinearly magnetized leads are of particular interest for a number of reasons. First, they show a nonequilibrium spin accumulation on the quantum dot due to spin-dependent tunneling. This spin accumulation has the tendency to block transport through the system and therefore gives rise to a tunnel magnetoresistance.~\cite{braig_rate_2005,rudziski_spin_2005,weymann_tunnel_2005,weymann_cotunneling_2005,hornberger_transport_2008} On the other hand, there is a precession of the dot spin due to an effective exchange field acting on the dot.~\cite{knig_interaction-driven_2003,braun_theory_2004,braun_hanle_2005,braun_frequency-dependent_2006} This exchange field arises due to spin-dependent virtual tunneling processes between the dot and the leads. It helps lifting the spin blockade by allowing the dot spin to precess out of its blocking position.
It is, thus, responsible for a deviation from the harmonic behavior of the conductance as a function of the angle between the magnetization directions of the leads.~\cite{knig_interaction-driven_2003} In the nonlinear transport regime, the interplay between spin accumulation and spin precession leads to a broad region of negative differential conductance.~\cite{braun_theory_2004} A more direct access to the exchange field is provided by the finite-frequency Fano factor that shows a resonance signal at the Larmor frequency associated with the exchange field.~\cite{braun_frequency-dependent_2006} A splitting of the Kondo resonance due to the exchange field was predicted using the numerical renormalization group~\cite{martinek_kondo_2003,martinek_kondo_2003-1,choi_kondo_2004,martinek_gate-controlled_2005,sindel_kondo_2007} and observed in recent experiments.~\cite{pasupathy_kondo_2004,hauptmann_electric-field-controlled_2008,hofstetter_ferromagnetic_2010} Recently, a new way to access the exchange field in a quantum-dot spin valve with an additional superconducting electrode was proposed.~\cite{sothmann_probing_2010}

Further theoretical studies addressed the current noise~\cite{buka_current_2000,cottet_positive_2004,weymann_theory_2007,weymann_shot_2008} and the full-counting statistics~\cite{lindebaum_spin-induced_2009} of single-level quantum dots coupled to ferromagnetic electrodes, as well as transport through multi-level dots,~\cite{weymann_cotunneling_2006,weymann_shot_2008} double dots,~\cite{weymann_spin-polarized_2007,hornberger_transport_2008,trocha_negative_2009} and carbon nanotube dots~\cite{koller_spin_2007,weymann_theory_2007,weymann_transport_2008,weymann_spin_2008,schenke_exchange_2009} coupled to ferromagnetic electrodes.

While the transport properties of quantum-dot spin valves have now been investigated in quite some detail, more extended models that include, e.g., the possibility to excite spin waves in the ferromagnetic electrodes have not yet been addressed. In this work, we consider the influence of spin waves on transport through a quantum-dot spin valve using a suitable extension of the real-time diagrammatic transport theory.~\cite{knig_zero-bias_1996,knig_resonant_1996,schoeller_transport_1997,knig_quantum_1999}
On the one hand we want to analyze the deviations from the idealized system. To this end, we study the modifications of the conductance which we find to be particularly pronounced for large polarizations of the leads. Here, the magnonic side peaks can exhibit negative differential conductance but can also surmount the ordinary conductance peaks depending on the magnetic configuration. We, furthermore, show that the excitation of spin waves can lead to an increased as well as to a decreased Fano factor. Additionally, we demonstrate how the magnonic modifications to the exchange field can be detected in the finite-frequency noise.
On the other hand, we want to address the question if the spin waves can generate completely new effects. To this end, we analyze the nonequilibrium distribution of the magnons which we find to be different for the source and drain electrode. Furthermore, we show how the magnons can drive a completely spin-polarized current without any external bias voltage.

The paper is organized as follows. In Sec.~\ref{sec:model} we introduce our model. The extension of the real-time diagrammatic technique to systems containing spin waves is presented in Sec.~\ref{sec:technique}. Our results are presented in Sec.~\ref{sec:results}. We conclude by giving a summary in Sec.~\ref{sec:conclusions}.

\section{\label{sec:model}Model}
\begin{figure}
    \includegraphics[width=0.48\textwidth]{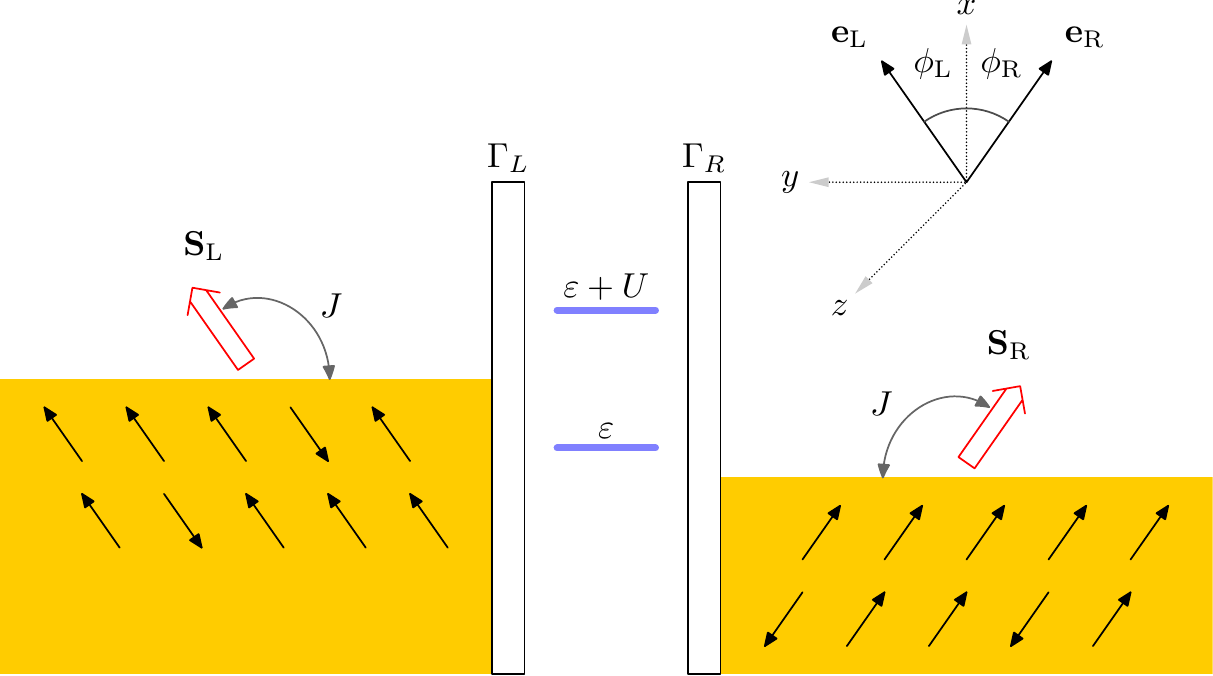}
    \caption{\label{fig:model}(Color online) Sketch of a quantum-dot spin valve with spin wave excitations in the leads and coordinate system used.}
\end{figure}

We consider transport through a single-level quantum dot weakly coupled to ferromagnetic leads with noncollinear magnetizations via tunneling barriers. In order to describe spin waves which may be excited in the leads, we model the lead magnetizations as macroscopically large spins localized in the respective lead as shown in Fig.~\ref{fig:model}.

The Hamiltonian of our system is the sum of five parts
\begin{equation}
	H=H_\text{dot}+H_r+H_\text{tun}+H_\text{spin}+H_\text{coupl}.
\end{equation}

The first term, $H_\text{dot}=\sum_{\sigma}\varepsilon c_\sigma^\dagger c_\sigma+Uc_\up^\dagger c_\up c_\down^\dagger c_\down$ describes the quantum dot in terms of a single, spin-degenerate level with energy $\varepsilon$ measured relative to the Fermi energies of the leads in equilibrium and Coulomb energy $U$ for double occupation.

The ferromagnetic leads are modeled as reservoirs of itinerant electrons, $H_r=\sum_{r\vec k\sigma=\pm}\varepsilon_{r\vec k}a_{r\vec k\sigma}^\dagger a_{r\vec k\sigma}$, $r=\text{L},\text{R}$. Here, $a_{r\vec k\sigma}^\dagger$ denotes the creation operator for electrons in lead $r$ with momentum $\vec k$ and spin $\sigma$ which have energy $\varepsilon_{r\vec k}$. The spin quantization axis of each lead is chosen parallel to the direction of its magnetization $\vec e_r$.

Due to the noncollinear geometry, it is convenient to quantize the spin on the dot in the direction perpendicular to the magnetizations of the leads, cf. the coordinate frame indicated in Fig.~\ref{fig:model}. In this case, the tunnel Hamiltonian describing the coupling between dot and leads is given by
\begin{multline}
H_\text{tun}=\sum_{r\vec k}\frac{t_r}{\sqrt{2}}\left[a_{r\vec k+}^\dagger \left(e^{i\phi_r/2}c_\up+e^{-i\phi_r/2}c_\down\right)\right.\\\left.
+a_{r\vec k-}^\dagger \left(-e^{i\phi_r/2}c_\up+e^{-i\phi_r/2}c_\down\right)\right]+\text{h.c.},
\end{multline}
where $\phi_r$ denotes the angle enclosed between the magnetization of lead $r$ and the $x$ axis. It should be noted that though individual terms violate spin conservation, the total tunnel Hamiltonian is spin conserving.

Using the Holstein-Primakoff representation, the localized spins are expressed in terms of bosonic operators
\begin{align}
	S_{rz}&=S-b^\dagger_rb_r,\label{eq:HP_1}\\
	S_{r+}&=\left(\sqrt{2S-b^\dagger_rb_r}\right)b_r,\label{eq:HP_2}\\
	S_{r-}&=b_r^\dagger\left(\sqrt{2S-b^\dagger_rb_r}\right).\label{eq:HP_3}
\end{align}
Since the magnetizations are macroscopic quantities, their fluctuations will be small and hence $b^\dagger_r b_r\ll S$. We, therefore, can restrict ourselves to the leading-order terms when expanding Eqs.~\eqref{eq:HP_1}-\eqref{eq:HP_3} in $1/S$. The spins are then treated as a bosonic degree of freedom with energy $\omega_b$, $H_\text{spin}=\sum_r\omega_b b_r^\dagger b_r$, corresponding to optical magnons with a single, dispersionless mode. The coupling between the localized spins and the spins of the itinerant electrons in the leads is given by
\begin{equation}\label{eq:coupl}
	\begin{split}
		H_\text{coupl}&=-J\sum_r\vec s_r\cdot \vec S_r\\&
		\approx-J\sum_{r\vec k}\left[S\left(a_{r\vec{k}+}^\dagger a_{r\vec{k}+}-a_{r\vec{k}-}^\dagger a_{r\vec{k}-}\right)\right.\\&\phantom{=}
		+\left.\sqrt{2S}\left(a_{r\vec{k}+}^\dagger a_{r\vec{k}-}b_r^\dagger+a_{r\vec{k}-}^\dagger a_{r\vec{k}+}b_r \right)\right].
	\end{split}
\end{equation}
The terms in the second line are independent of the magnon number.
They can be absorbed into the part of the Hamiltonian describing the lead electrons, yielding spin-dependent electron energies, $\varepsilon_{r\vec k\sigma}=\varepsilon_{r\vec k}-\sigma JS$. As a consequence, the densities of states will also acquire a spin dependence, $\rho_{r,\sigma}(\omega)$. In the following, we will refer to the electrons with larger (smaller) density of states at the Fermi energy as majority, $\sigma=+$ (minority, $\sigma=-$) spin electrons. Since in transport only a small energy window at the Fermi energy is relevant, we will take the densities of state to be independent of energy, $\rho_{r,\sigma}=\rho_{r,\sigma}(E_F)$.  In this case, the tunnel matrix elements $t_r$ can be related to the tunneling rate $\Gamma_{r\pm}$ for a majority/minority spin electron from lead $r$ via $\Gamma_{r\pm}=2\pi|t_r|^2\rho_{r,\pm}$. We furthermore define the average tunneling rate $\Gamma_r=(\Gamma_{r+}+\Gamma_{r-})/2$.

The asymmetry between majority and minority spins is then characterized by the spin polarization $p_r=(\rho_{r,+}-\rho_{r,-})/(\rho_{r,+}+\rho_{r,-})$.
We note that in general for a non-constant density of states, the total number of majority spin electrons may be larger or smaller than the total number of minority spin electrons, depending on the precise form of the band structure, the size of the splitting and the position of the Fermi energy. Hence, the localized spin may point parallel or antiparallel to the majority spin direction corresponding to positive or negative values of the polarization $p_r$. We note that a nonconstant density of states gives rise to a modified exchange field, Eq.~\eqref{eq:exchangefield}, see below.
 
The terms in the last line of Eq.~\eqref{eq:coupl} describe spin-flip interactions between lead electrons and the magnons. In order to remove this interaction, we apply to the Hamiltonian the canonical transformation~\cite{nagaev_magnon_1998} $\tilde H=e^A H e^{-A}$ with generator
\begin{equation}
	A=-\sum_{r\vec k}\lambda\left(a_{r\vec{k}+}^\dagger a_{r\vec{k}-}b_r^\dagger-a_{r\vec{k}-}^\dagger a_{r\vec{k}+}b_r \right)
\end{equation}
where $\lambda=J\sqrt{2S}/(\omega_b+\varepsilon_{r\vec{k}+}-\varepsilon_{r\vec{k}-})=J\sqrt{2S}/(\omega_b-2JS)$.
Neglecting terms of order $\lambda^2$ that can be absorbed into the energies $\varepsilon_{r\vec k\sigma}$, the transformed Hamiltonian takes the form
\begin{equation}
	\tilde H=H_\text{dot}+H_r+H_\text{spin}+\tilde H_\text{tun}
\end{equation}
with the transformed tunneling Hamiltonian
\begin{equation}\label{eq:Htuntransformed}
	\tilde H_\text{tun}=\sum_{r\vec k}\frac{t_r}{\sqrt{2}}\left(a_{r\vec k+}^\dagger \tilde c_{r\up}+a_{r\vec k-}^\dagger \tilde c_{r\down}\right)+\text{h.c.}
\end{equation}
where
\begin{align}
	\tilde c_{r\up}=&\left(1-\frac{\lambda^2}{2}b_r^\dagger b_r\right)\left(e^{i\phi_r/2}c_\up+e^{-i\phi_r/2}c_\down\vphantom{\frac{1}{2}}\right)
	\label{eq:newdotoperators1}\\
	&-\lambda b_r^\dagger\left(-e^{i\phi_r/2}c_\up+e^{-i\phi_r/2}c_\down\vphantom{\frac{1}{2}}\right),\notag\\
	\tilde c_{r\down}=&\left(1-\frac{\lambda^2}{2}b_r^\dagger b_r\right)\left(-e^{i\phi_r/2}c_\up+e^{-i\phi_r/2}c_\down\vphantom{\frac{1}{2}}\right)\label{eq:newdotoperators2}\\
	&+\lambda b_r\left(e^{i\phi_r/2}c_\up+e^{-i\phi_r/2}c_\down\vphantom{\frac{1}{2}}\right).\notag
\end{align}
In writing down the transformed tunnel Hamiltonian, Eq.~\eqref{eq:Htuntransformed}, we neglected terms of order $\lambda^2$ that do not involve bosonic operators as these terms do not yield a contribution to order $\lambda^2$ in the diagrammatic expansion. The canonical transformation gives rise to new processes in the tunnel Hamiltonian. Apart from tunneling events that are already present for the ordinary quantum-dot spin valve,~\cite{braun_theory_2004} we encounter additional terms in which a magnon is emitted/absorbed and the spin of the tunneling electron is flipped. Hence, the total tunnel Hamiltonian does not conserve the electron spin any longer but just the sum of electron spin and angular momentum of the magnons.

\section{\label{sec:technique}Real-time diagrammatic technique}
In order to investigate the transport properties of our system, we extend the real-time diagrammatic technique developed in Refs.~\onlinecite{knig_zero-bias_1996,knig_resonant_1996,schoeller_transport_1997,knig_quantum_1999} and adapted to systems with ferromagnetic leads in Refs.~\onlinecite{knig_interaction-driven_2003}~and~\onlinecite{braun_theory_2004} to include the spin-wave degrees of freedom.

\subsection{\label{ssec:masterequation}Reduced density matrix and master equation}
The basic idea of the real-time diagrammatic technique is to integrate out the non-interacting, fermionic degrees of freedom of the leads in order to obtain an effective description of the reduced system which is characterized by the state of the quantum dot and the number of magnons in the left and right lead. The quantum dot can be either empty, occupied with a spin up or a spin down electron, or doubly occupied. We denote these states as $\ket{0}$, $\ket{\up}$, $\ket{\down}$, and $\ket{d}$ with  energies $E_0=0$, $E_\up=E_\down=\varepsilon$, and $E_d=2\varepsilon+U$, respectively. The number of magnons is characterized by $\ket{\vec n}=\ket{n_\text{L},n_\text{R}}$. The total energy of a state $|\xi\rangle=|\chi,\vec n\rangle$ is then given by $E_\xi=E_\chi+(n_\text{L}+n_\text{R})\omega_b$. In order to keep the dimension of the Hilbert space finite, we introduce a maximal magnon number $N_\text{max}$ for actual computations and check that our results are independent of the cut-off value. For the parameters chosen in the analysis below, it turned out that it is sufficient to take into account at most four magnons in each lead.

After tracing out the fermionic degrees of freedom in the leads, the system is described by a reduced density matrix $\rho^\text{red}$ with matrix elements $P_{\xi_1}^{\xi_2}=\langle\xi_2|\rho^\text{red}|{\xi_1}\rangle$. For the diagonal matrix elements we introduce the abbreviation $P_\xi=P_\xi^\xi$.

In the stationary state, the density matrix elements obey a generalized master equation of the form
\begin{equation}\label{eq:master}
	0=\dot P_{\xi_1}^{\xi_2}=i(E_{\xi_2}-E_{\xi_1})P_{\xi_1}^{\xi_2}+\sum_{\xi_1'\xi_2'}W_{\xi_1\xi_1'}^{\xi_2\xi_2'}P_{\xi_1'}^{\xi_2'}.
\end{equation}
Here, the kernels $W_{\xi_1\xi'_1}^{\xi_2\xi'_2}$ describe transitions due to the dot-lead coupling. Similarly as for the diagonal density matrix elements, we introduce the short-hand form $W_{\xi\xi'}^{\xi\xi'}=W_{\xi\xi'}$. The kernels that enter the master equation are defined as irreducible self-energy diagrams for the dot propagator on the Keldysh contour and can be expanded perturbatively in the tunnel coupling strength $\Gamma$. In the following, we will restrict ourselves to terms that are of first order in $\Gamma$. The diagrammatic rules necessary for the evaluation of the diagrams are summarized in Appendix~\ref{app:diagrams}.

In the following, we will always assume the magnon energies to be much larger than the tunnel coupling, $\omega_b\gg\Gamma$. In this case, the master equation for the matrix elements which are off-diagonal in the magnon number becomes $i(E_{\xi_2}-E_{\xi_1})P_{\xi_1}^{\xi_2}=0$ to first order in the tunnel coupling. This implies that only matrix elements diagonal in the boson number have to be taken into account.
In this case, the reduced density matrix takes a block-diagonal form $\rho^\text{red}=\rho_\text{dot}\otimes\rho_\text{boson}$, where each block can be written as
\begin{equation*}
	\left(\begin{array}{cccc}
	P_{0\vec n}	& 0 			& 0			& 0 \\
	0		& P_{\up\vec n}		& P_{\down\vec n}^{\up}	& 0 \\
	0		& P_{\up\vec n}^{\down}	& P_{\up\vec n}		& 0 \\
	0		& 0			& 0			& P_{d\vec n}
	\end{array}\right).
\end{equation*}
To allow for a compact notation, we write the density matrix as a vector consisting of blocks of the form $(P_{0\vec n},P_{\up\vec n},P_{\down\vec n},P_{d\vec n},P^{\up}_{\down\vec n},P^{\down}_{\up\vec n})^T$. The normalization of the reduced density matrix can then be cast into the form $\mathbf e^T\rho^\text{red}=1$, where $\mathbf e^T$ is a vector consisting of blocks of the form $(1,1,1,1,0,0)$.

To allow an easier physical interpretation of the density matrix elements, it is convenient to express them  in terms of the average dot occupations $P_{0\vec n}$, $P_{1\vec n}=P_{\up\vec n}+P_{\down\vec n}$ and $P_{d\vec n}$ and the average spin on the quantum dot
\begin{equation}
	S_{x\vec n}=\frac{P_{\down\vec n}^\up+P_{\up\vec n}^\down}{2},\; S_{y\vec n}=i\frac{P_{\down\vec n}^\up-P_{\up\vec n}^\down}{2},\; S_{z\vec n}=\frac{P_{\up\vec n}-P_{\down\vec n}}{2}
\end{equation}
in the presence of $\vec n$ magnons. The set of master equations can then be split into one describing the average charge and one describing the average spin on the dot.

The master equation for the occupation probabilities is given by
\begin{equation}\label{eq:mastercharge}
	\frac{d}{dt}
	P_{\chi\vec n}=\sum_r{\sum_{\vec m}}\left(\sum_{\chi'}M_{\chi\vec n,\chi'\vec m}^{(r)}P_{\chi'\vec m}+V_{\chi\vec m}^{(r)}\vec S_{\vec m}\cdot \vec e_r\right).
\end{equation}
Here, $M_{\chi\vec n,\chi'\vec m}^{(r)}$ denotes transition rates from state $\chi'\vec m$ to state $\chi \vec n$, while $V_{\chi\vec m}^{(r)}$ characterizes the dependence of the dot occupation on the accumulated spin. Their precise form is given in Appendix~\ref{app:mastereq}. The sum over $\vec m$ only gives contributions if in a process involving the left (right) lead the number of magnons in the right (left) lead is kept fixed while the number of magnons in the left (right) lead changes by at most one due to the conservation of angular momentum in each tunneling event.

This restriction in the number of excited magnons is different from the case of a quantum dot coupled to a vibrational degree of freedom. In the latter case, the number of phonons that can be excited in a tunneling event is limited only by the applied bias voltage while the number of phonons that can be absorbed is only limited by the number of excited phonons,~\cite{koch_current-induced_2006} giving rise to a large number of conductance sidebands~\cite{mitra_phonon_2004,koch_franck-condon_2005,koch_theory_2006} as well as to a phonon distribution width that is nonperturbative in the electron-phonon coupling.~\cite{koch_current-induced_2006} Furthermore, the transition rates in the vibrational case show a nontrivial dependence on the initial and final number through the Franck-Condon factors that influence the transport properties crucially, e.g., by leading to a suppression of transport for small bias voltages.~\cite{koch_franck-condon_2005,koch_theory_2006}

The dot spin obeys the Bloch-type equation
\begin{equation}\label{eq:masterspin}
	\frac{d\vec S_{\vec n}}{dt}=\left(\frac{d\vec S_{\vec n}}{dt}\right)_\text{acc}+\left(\frac{d\vec S_{\vec n}}{dt}\right)_\text{rel}+\left(\frac{d\vec S_{\vec n}}{dt}\right)_\text{prec}
\end{equation}
where
\begin{equation}\label{eq:masterspinaccumulation}
	\left(\frac{d\vec S_n}{dt}\right)_\text{acc}=\sum_r{\sum_{\vec m}}\sum_\chi F_{\chi\vec n\vec m}^{(r)}P_{\chi\vec m} \vec e_r,
\end{equation}
\begin{equation}\label{eq:masterspinrelaxation}
	\left(\frac{d\vec S_n}{dt}\right)_\text{rel}=-\sum_{r}G^{(r)}\vec S_n,
\end{equation}
\begin{equation}\label{eq:masterspinprecession}
	\left(\frac{d\vec S_{\vec n}}{dt}\right)_\text{prec}=\vec S_{\vec n}\times\sum_r\vec B_{\vec n}^{(r)}.
\end{equation}
The dynamics of the dot spin is governed by three terms. The first one, Eq.~\eqref{eq:masterspinaccumulation}, which depends only on the occupation probabilities, describes the accumulation of spin on the dot due to electrons tunneling onto the empty dot or electrons leaving the doubly occupied dot. Here, the sum over $\vec m$ is subject to the same restriction as in the master equation for the occupation probabilities. The second term, Eq.~\eqref{eq:masterspinrelaxation}, which is proportional to the accumulated spin, describes the decay of the dot spin due to tunneling out of electrons or tunneling in with a spin opposite to the dot spin, forming a spin singlet on the dot. The precise form of the functions $F_{\chi\vec n\vec m}^{(r)}$ and $G^{(r)}$ that enter the accumulation and relaxation term is given in Appendix~\ref{app:mastereq}. Finally, the third term, Eq.~\eqref{eq:masterspinprecession}, describes the precession in the exchange field generated by virtual tunneling between the dot and the leads. It is given by
\begin{figure}
	\includegraphics[width=\columnwidth]{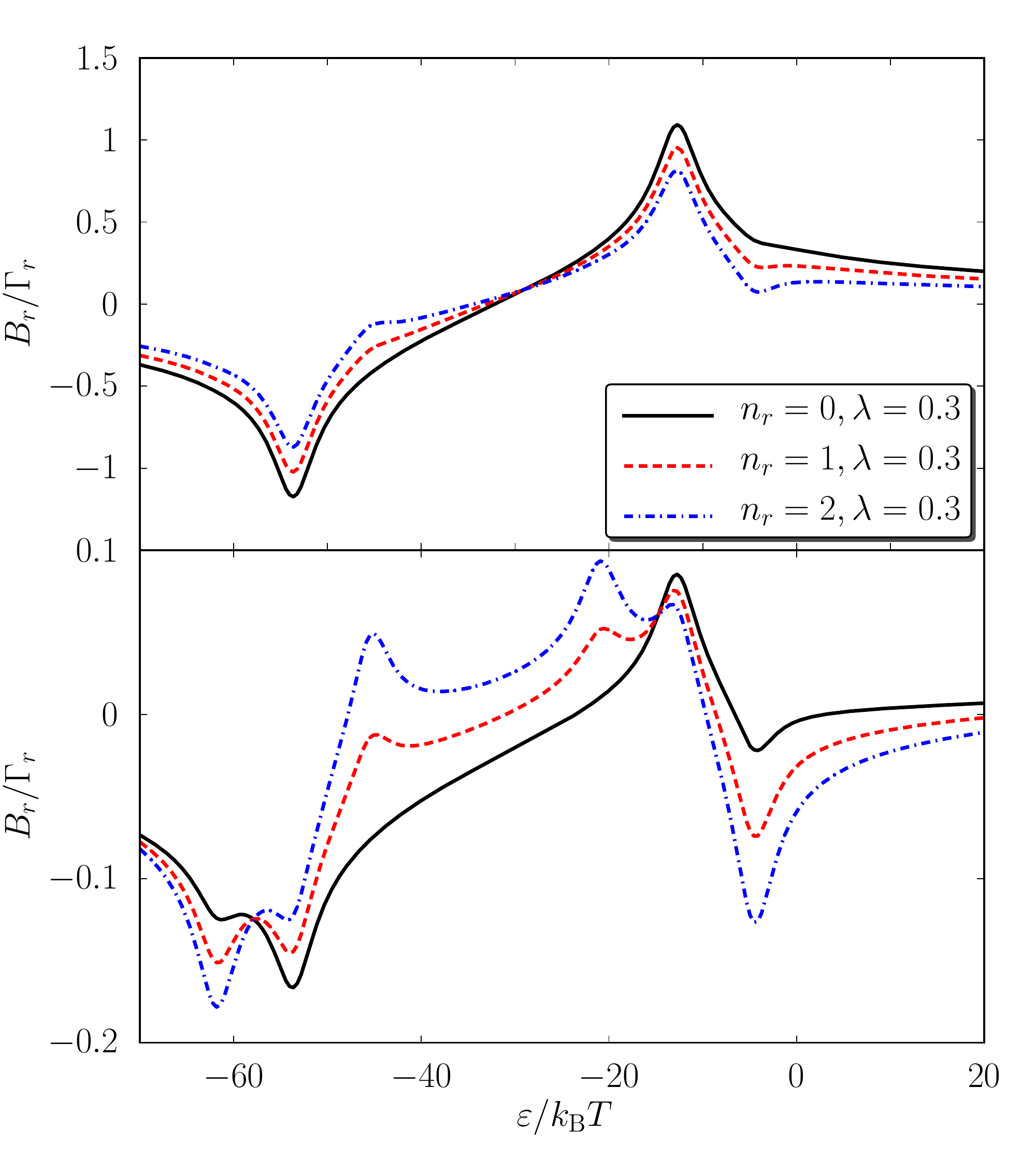}
	\caption{\label{fig:exchangefield}(Color online) Exchange field as a function of the level position $\varepsilon$ for $p_r=0.9$ (upper panel) and $p_r=0.1$ (lower panel). Other parameters are $U=50\kB T$, $\omega_B=10\kB T$.}
\end{figure}
\begin{widetext}
\begin{multline}\label{eq:exchangefield}
	\vec B_{\vec n}^{(r)}=-\vec e_r\frac{\Gamma_r}{\pi}\left\{(1-\lambda^2n_r)p_r\left[\Phi_r(\varepsilon)-\Phi_r(\varepsilon+U)\right]
	+\frac{\lambda^2}{2}\left[2\ln\frac{\beta W}{2\pi}-(1+p_r)(1+n_r)\Phi_r(\varepsilon-\omega_b)\right.\right.\\\left.\left.
	+(1-p_r)n_r\Phi_r(\varepsilon+\omega_b)+(1+p_r)n_r\Phi_r(\varepsilon+U-\omega_b)-(1-p_r)(1+n_r)\Phi_r(\varepsilon+U+\omega_b)\vphantom{\frac{1}{2}}\right]\right\},
\end{multline}
\end{widetext}
where $\Phi_r(x)=\RPG{(x-\mu_r)}$, and $\Psi$ is the digamma function. Compared to the ordinary quantum-dot spin valve,~\cite{braun_theory_2004} the exchange field contains new terms proportional to $\lambda^2$ which arise form virtual tunneling processes that emit or absorb magnons in the intermediate state. Furthermore, the terms already present in the absence of spin waves experience a magnon-number-dependent renormalization $1-\lambda^2n_r$ that has the tendency to reduce the strength of the exchange field. In Fig.~\ref{fig:exchangefield} we show the exchange field as a function of the level position. The new terms give rise to side peaks and dips. As the original exchange field is rather small for small polarizations, the magnonic features tend to be more pronounced for small polarizations.

The logarithmic divergency in Eq.~\eqref{eq:exchangefield} that is cut-off by the bandwidth of the lead electrons $W$ (in the following, we assume $W=100\kB T$) arises as the rate for emitting and absorbing a magnon differ from each other. If the system is, e.g., in a state without magnons and the dot is singly occupied, only a spin down electron can leave the dot to the leads by a magnon-assisted process while only a spin up electron can enter the dot in such a process. Hence, only the energy of the spin-down state is renormalized by the magnonic processes, thereby giving rise to a diverging energy shift between spin up and spin down electrons and hence to a diverging exchange field.

In order to take into account the finite life time of spin waves, e.g., due to scattering from phonons or electrons or due to magnon-magnon interactions, we include phenomenological relaxation terms $-\frac{1}{\tau}\left(P_{\chi\vec n}-P^\text{eq}_{\vec n}\sum_{\vec n}P_{\chi\vec n}\right)$ and $-\frac{1}{\tau}\left(\vec S_{\chi\vec n}-P^\text{eq}_{\vec n}\sum_{\vec n}\vec S_{\chi\vec n}\right)$ into the master Eqs.~\eqref{eq:mastercharge} and~\eqref{eq:masterspin}. These terms describes a relaxation towards the equilibrium distribution of magnons $P_n^\text{eq}=e^{-n\omega_b/\kB T_B}/(1-e^{-\omega_b/\kB T_B})$ on a time scale $\tau$. We allow for the most general case of a magnon temperature $T_B$ that differs from the electron temperature $T$.

Besides making our model more realistic, the relaxation terms also ensure that the magnon number in the drain lead remains finite. While the emission of a magnon when tunneling out of the dot transfers the electron from minority to majority spin and thus gives rise to a rate proportional to $1+p_r$, the absorption flips the spin in the opposite direction and therefore yields a rate proportional to $1-p_r$. Hence, without any relaxation mechanism, the number of magnons would grow without bound in the drain lead.

\subsection{\label{ssec:current}Current and current noise}
We define the current through the tunnel barrier $r$ as the change in the number of electrons in lead $r$ multiplied with the electron charge, $\hat I_r=-ie[H,\sum_{\vec k\sigma}a_{r\vec k\sigma}^\dagger a_{r\vec k\sigma}]$. While in the stationary state the current at zero frequency obeys $\langle \hat I\rangle=\langle \hat I_\text{L}\rangle=-\langle \hat I_\text{R}\rangle$ due to current conservation, at finite frequencies the current measured in the source-drain circuit in general differs from the currents through the tunnel barriers due to occurrence of displacement currents. These can be taken into account according to the Ramo-Shockley theorem~\cite{Shockley_currents_1938,Ramo_currents_1939,blanter_shot_2000} by considering the total current as the sum of the current $\hat I_r$ through each tunnel barrier weighted with the corresponding capacitance $C_r$ of the barrier, $\hat I=(C_\text{L}\hat I_\text{L}+C_\text{R}\hat I_\text{R})/(C_\text{L}+C_\text{R})$. As the capacitances of the tunnel barriers are much less sensitive to the geometry than the tunnel couplings, in the following, we assume symmetric capacitances while allowing for asymmetric tunnel couplings. We therefore have $\hat I=(\hat I_\text{L}+\hat I_\text{R})/2$. Diagrammatically, the current can be evaluated as
\begin{equation}\label{eq:current}
	\langle\hat I\rangle
	=\frac{e}{2\hbar} \mathbf e^T \mathbf W^I \mathbf P,
\end{equation}
where $\vec P$ denotes the vector of density matrix elements and
$\mathbf W^I$ are kernels in which one internal vertex is replaced by a current vertex. Since the current operator equals the tunnel Hamiltonian apart from constant factors, these replacements only give rise to factors of $\pm 1$ depending on the position of the vertex on the contour and whether the electron tunnels in or out of the dot (cf. Appendix~\ref{app:diagrams} for details).

We define the frequency-dependent current noise~\cite{braun_frequency-dependent_2006} as the Fourier transform of $S(t)=\langle\hat I(t)\hat I(0)\rangle+\langle\hat I(0)\hat I(t)\rangle-2\langle\hat I\rangle$,
\begin{multline}\label{eq:noise}
	S(\omega)=\int_0^\infty dt \left(\langle\hat I(t)\hat I(0)\rangle+\langle\hat I(0)\hat I(t)\rangle\right)\left(e^{i\omega t}+e^{-i\omega t}\right)\\
	-4\pi\delta(\omega)\langle \hat I\rangle.
\end{multline}
By choosing a symmetrized expression for the current noise, we restrict ourselves to real noise measurable in a classical detector~\cite{landau_statistical_1958} in contrast to the unsymmetrized expression which is complex and describes detector-dependent emission and absorption processes.~\cite{lesovik_detection_1997,aguado_double_2000,gavish_detection_2000,kck_full-frequency_2003,engel_asymmetric_2004,schoelkopf_qubits_2003}

Following Ref.~\onlinecite{braun_frequency-dependent_2006}, we evaluate the finite-frequency noise diagrammatically to first order in the tunnel coupling as
\begin{multline}\label{eq:diagrammaticnoise}
S(\omega)=\frac{e^2}{2\hbar}\vec e^T\left[\vec W^{II}+\vec W^I\left(\boldsymbol\Pi_0^{-1}(\omega)-\vec W\right)\vec W^I\right]\vec P\\-2\pi\delta(\omega)\langle \hat I\rangle^2+(\omega\to-\omega).
\end{multline}
Here, the matrices $\mathbf W^{II}$ correspond to self-energies with two tunnel vertices replaced by current vertices. Furthermore, we introduced the free propagator of the quantum dot,
\begin{equation}
	\Pi_0(\omega)^{\chi_1\chi'_1}_{\chi_2\chi'_2}=\frac{i\delta_{\chi_1\chi'_1}\delta_{\chi_2\chi'_2}}{\varepsilon_{\chi_2}-\varepsilon_{\chi_1}-\omega+i0^+}.
\end{equation}

\section{\label{sec:results}Results}
In this section, we discuss our results for the current, conductance and current noise of the quantum-dot spin valve in the presence of spin waves. Unless stated otherwise, we assume a symmetric system with equal polarizations for both leads, $p_\text{L}=p_\text{R}\equiv p$, and equal tunnel couplings, $\Gamma_\text{L}=\Gamma_\text{R}\equiv\Gamma/2$. Furthermore, we assume that the bias voltage is applied symmetrically, $V_\text{L}=-V_\text{R}=V/2$, too.

\subsection{\label{ssec:processes}Transport processes}
\begin{figure}
	\includegraphics[width=\columnwidth]{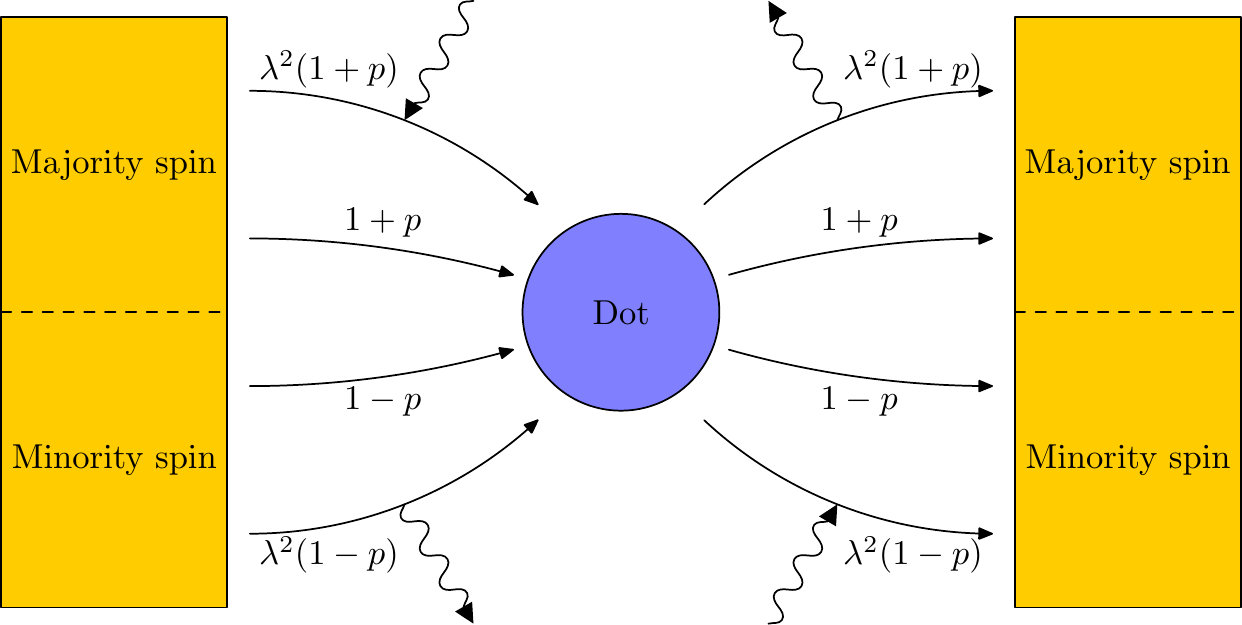}
	\caption{\label{fig:processes}(Color online) Transport processes that arise in a quantum-dot spin valve with spin wave degrees of freedom in the leads as well as their dependence on the polarization $p$.}
\end{figure}
Before discussing the transport properties in detail, we give an overview of the different transport processes that can occur in the presence of spin waves in the leads. In Fig.~\ref{fig:processes}, we show all transport processes that can arise. As for the ordinary quantum-dot spin valve, majority (minority) spin electrons can tunnel from the source onto the dot as well as from the dot into the drain lead. The corresponding rates are given by $(1\pm p)\Gamma/4$.

Additionally, we now have processes that involve the emission/absorption of magnons. For example, a majority spin from the source can flip its spin by absorbing a magnon to become a minority spin that ends up on the quantum dot. Hence, the rate for this process is given by $\lambda^2n_\text{L}(1+p)\Gamma/4$. Similarly, a minority spin electron can emit a magnon to become a majority spin that tunnels onto the dot. Here, the corresponding rate is given by $\lambda^2(1+n_\text{L})(1-p)\Gamma/4$.

Furthermore, there are two processes involving the magnons in the drain lead. Now, a minority spin of the drain can leave to the dot, emit a magnon and end up as a majority spin. The rate for this process is given by $\lambda^2(1+n_\text{R})(1+p)\Gamma/4$. The opposite process which starts with a majority spin, absorbs a magnon and ends up as a minority spin finally has a rate given by $\lambda^2n_\text{R}(1-p)\Gamma/4$.

Hence, we find that for a given spin direction on the dot the magnonic processes have the opposite dependence on the polarization as the normal processes. We will show in the following how this unconventional polarization dependence gives rise to a number of interesting transport properties.

\subsection{\label{ssec:cond}Magnon-assisted tunneling}
\begin{figure}
	\includegraphics[width=\columnwidth]{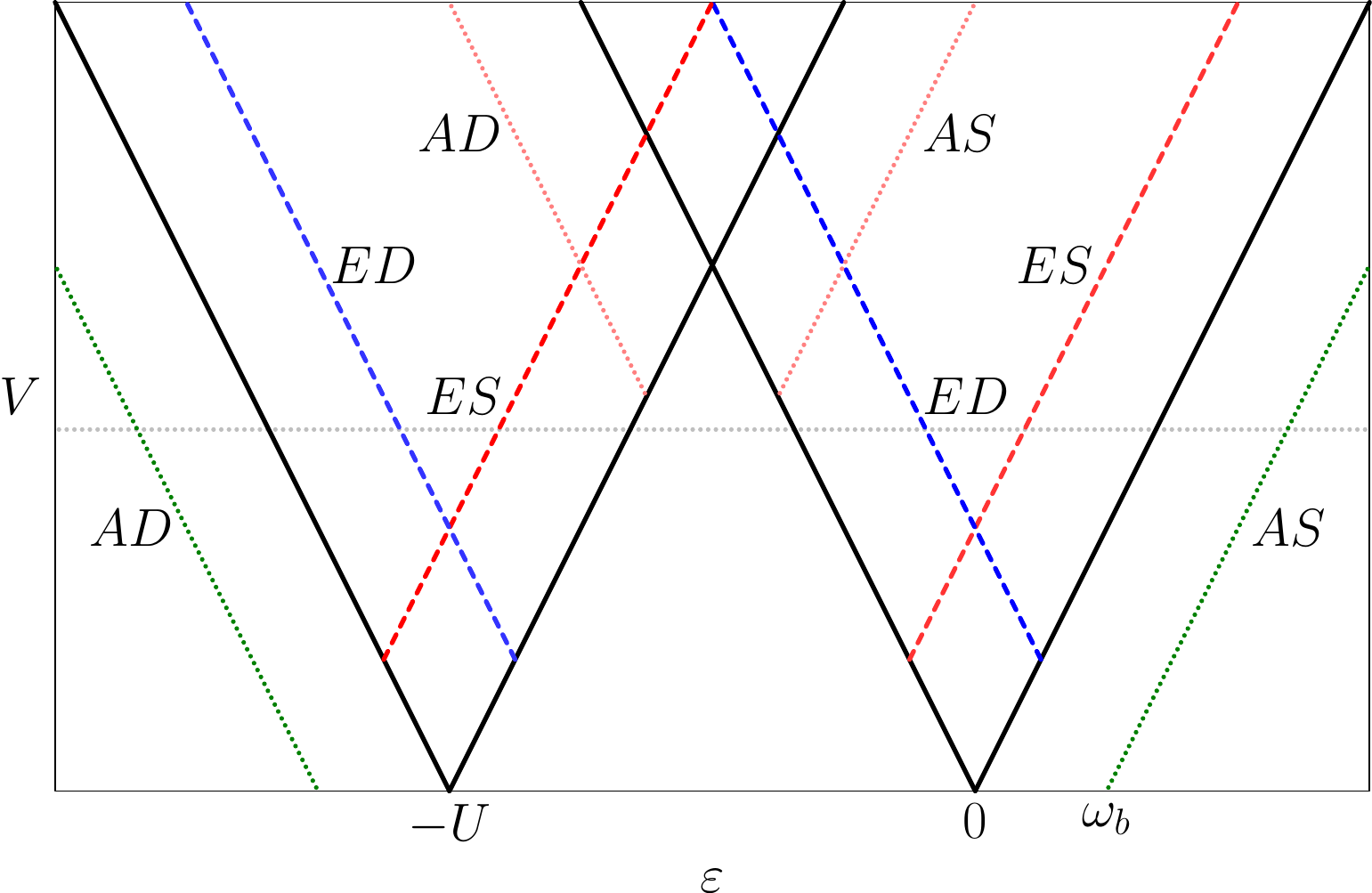}
	\caption{\label{fig:conductance}(Color online) Schematic of the differential conductance in the $\varepsilon-V$ plane. The labels at the conductance lines indicate whether a magnon is emitted (E) or absorbed (A) in the source (S) or drain (D) electrode. The dotted horizontal line marks the position of the cuts shown in Fig.~\ref{fig:conductancecut}.}
\end{figure}
We now turn to the discussion of the differential conductance $G=dI/dV$ for arbitrary bias and gate voltages. In Fig.~\ref{fig:conductance}, we schematically show the conductance for collinear magnetizations~\footnote{}.
In addition to the resonances of the dot level with the Fermi energy of the left and right lead which are marked by the thick black lines, a number of sidebands related to the emission and absorption of magnons occur. At the red (light) dashed lines labeled $ES$, the tunneling electrons become able to emit magnons in the source lead, while at the blue (dark) dashed lines $ED$ the emission of magnons in the drain lead becomes energetically possible. The dashed green (dark) lines indicate where transport through the dot becomes possible by absorbing magnons to enter the empty dot with both level above the two Fermi energies ($AS$) or to leave the doubly occupied dot with both levels below the two Fermi energies ($AD$). Furthermore, there are additional peaks marked by the orange (light) dotted lines $AS$ and $AD$ at which it becomes possible to absorb a magnon in order to enter the doubly occupied state which otherwise would have been out of reach energetically. Conductance sidebands inside the Coulomb blockade regime are absent because there a spin accumulates on the dot in such a way as to suppress all magnonic processes.

As discussed above, the different magnonic processes depend on the polarization of the leads as $1+p$ or $1-p$ depending on whether they involve a majority or minority spin electron in the lead. As a consequence, the different sidebands will have different strengths. In particular, while line $ES$ is suppressed with increasing the polarization, line $ED$ is increased. Hence, the conductance map no longer exhibits particle-hole symmetry $\varepsilon\to U-\varepsilon$ due to the presence of the magnons.

As discussed in Sec.~\ref{sec:model} the polarization of the leads can take either sign: a positive (negative) sign indicates that the majority spins of the carriers at the Fermi energy are parallel (antiparallel) to the direction of the macroscopic magnetic moment.
We now discuss how the conductance plot changes for the different combinations of positive and negative polarizations in the two leads. The case of two negative polarizations is related to the case of two positive ones by the transformation $\varepsilon\to U-\varepsilon$. This maps all lines involving the emission of magnons onto lines involving the absorption of magnons and vice versa and therefore exchanges the factors $1+p$ and $1-p$ in the magnonic rates. For polarizations of opposite signs,  the conductance plot becomes particle-hole symmetric. However, now the strength of the magnonic side bands depends on the direction of current flow. While they are suppressed for one direction by $1-p$ they are enhanced by $1+p$ in the other direction. In the rest of this work, we will restrict ourselves to the case in which both polarizations are positive. All other cases can be related to this one by the above symmetry considerations.

\begin{figure}
	\includegraphics[width=\columnwidth]{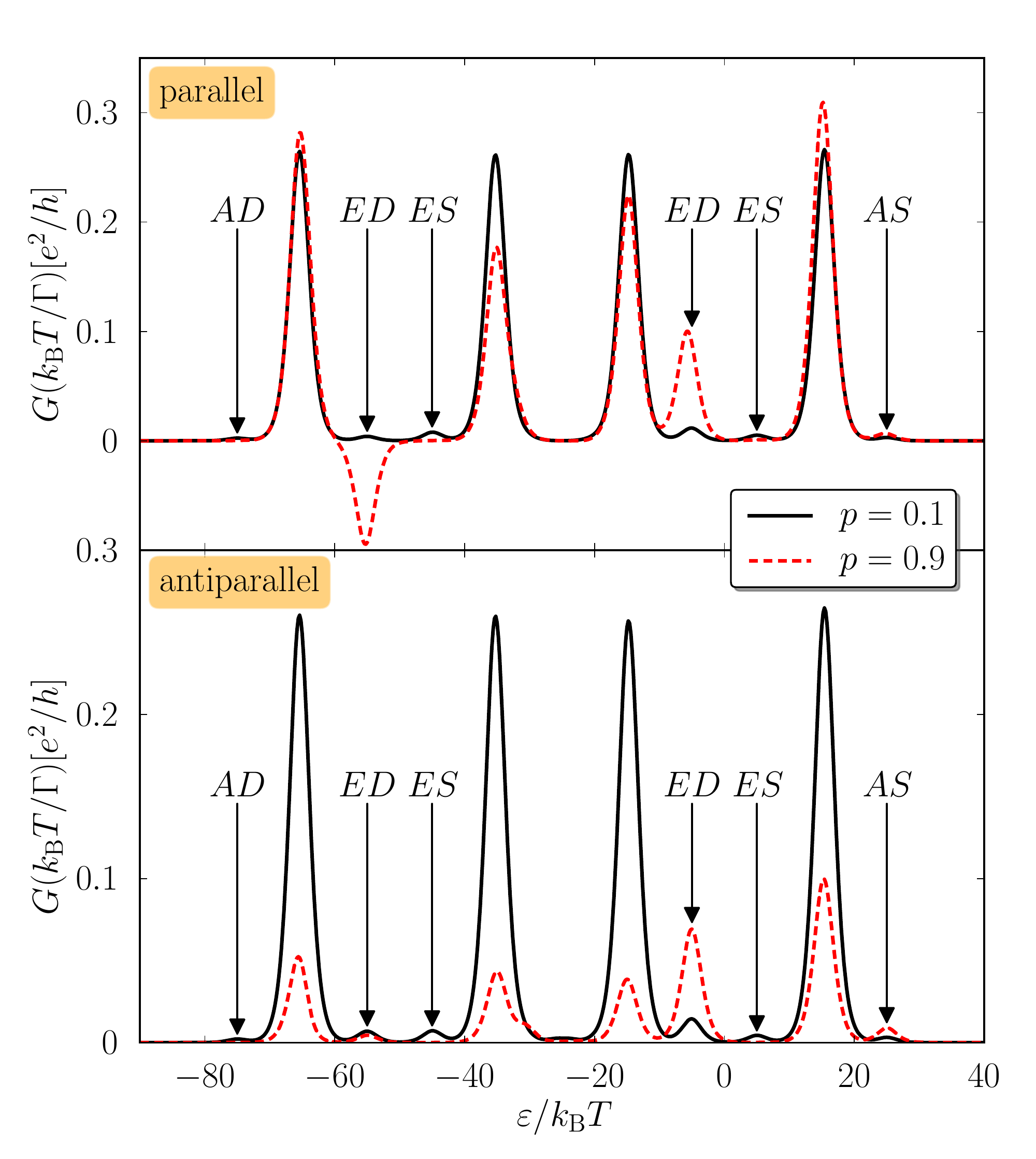}
	\caption{\label{fig:conductancecut}(Color online) Differential conductance as a function of the level position for $V=30\kB T$, $U=50\kB T$, $\omega_B=10\kB T$, $\Gamma_\text{L}=\Gamma_\text{R}$, $\tau=2/\Gamma$, $\lambda=0.3$, $T_B=5T$ for small and large polarizations. In the upper panel, we show the case of parallel magnetizations, while in the lower panel they are chosen to be antiparallel. As in Fig.~\ref{fig:conductance}, the labels of the side peaks indicate the emission (E) and absorption(E) of magnons in source (S) and drain (D) lead.}
\end{figure}
We now turn to the discussion of the conductance at fixed bias voltage shown in Fig.~\ref{fig:conductancecut}, corresponding to a cut along the dotted horizontal line in Fig.~\ref{fig:conductance}. In the upper panel of Fig.~\ref{fig:conductancecut}, we show the differential conductance for parallel magnetizations as a function of the level position for a given bias voltage. For small polarizations, there are four large peaks at which the dot levels are in resonance with either the left or right Fermi energy, corresponding to the thick black lines in Fig.~\ref{fig:conductance}. These are accompanied by much smaller conductance peaks at distances $\pm\omega_b$. From left to right, these correspond to lines $AD$, $ED$, $ES$, $ED$, $ES$, and $AS$ in Fig.~\ref{fig:conductance}. The lines corresponding to the absorption of magnons ($AS$ and $AD$) are only present for sufficiently high magnon temperatures, otherwise the number of magnons in the leads is too small to make these side peaks visible.

For large polarizations, there are two new effects. On the one hand, the two side peaks associated with lines labeled $ES$ are suppressed by $1-p$ and become invisible. On the other hand, we find that the side peak corresponding to left line $ED$ in Fig.~\ref{fig:conductance} shows a strong negative differential conductance. This is due to the formation of a trapping state. A spin down electron can leave the doubly occupied dot by exciting a magnon in the drain lead leaving the dot in the spin up state. This blocks further transport as tunneling in of an electron is suppressed by the small density of states for spin down electrons in the source lead.

We now turn to the case of antiparallel magnetizations. For small polarizations, the conductance resembles the one for the parallel configuration because the effects of the finite polarization are only weak. For large polarizations, there is a large spin accumulation on the dot which is antiparallel to the magnetization of the drain lead. When transport takes place through the states $\ket{0}$ and $\ket{\sigma}$ only, the rate for normal tunneling into the drain is suppressed by $1-p$ while the rate for tunneling via a spin flip is proportional to $\lambda^2(1+p)$. Hence, for large polarizations and not too small couplings to the magnons, the side peaks can dominate over the main peaks. On the other hand, for transport through the states $\ket{\sigma}$ and $\ket{d}$, the bottleneck is given by the tunneling in of electron. This bottleneck cannot be overcome by magnonic processes. Hence, we do not find large side peaks in this case.

In summary, this means that for large polarizations the magnons tend to decrease the current for transport through singly and doubly occupied dot in the parallel configuration while they tend to increase the current for transport through the empty and singly occupied dot for antiparallel magnetizations. As a consequence, the tunnel magnetoresistance will be reduced by the magnons for all transport regimes.

\subsection{\label{ssec:magnondistribution}Nonequilibrium magnon distribution}
\begin{figure}
	\includegraphics[width=\columnwidth]{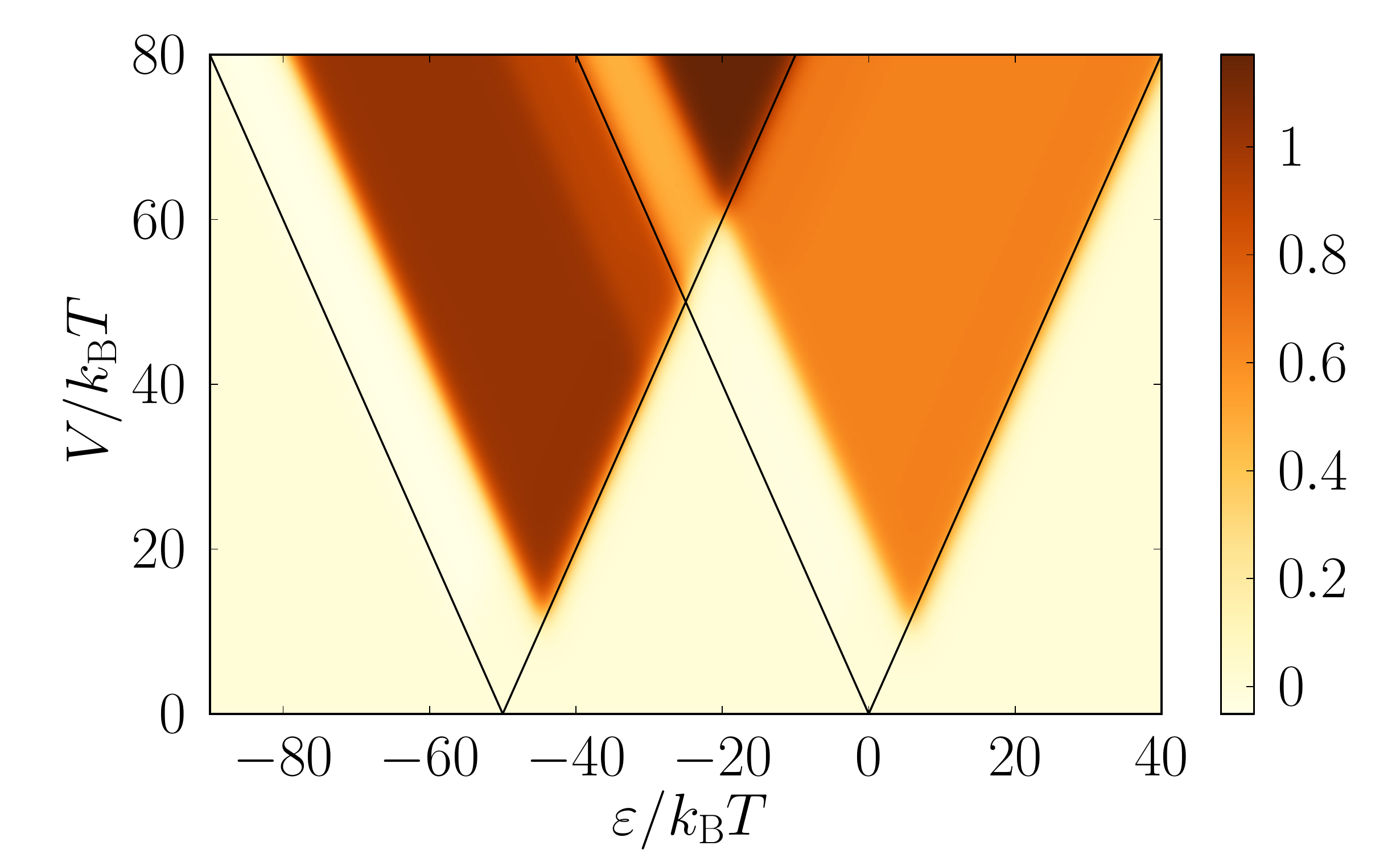}
	\includegraphics[width=\columnwidth]{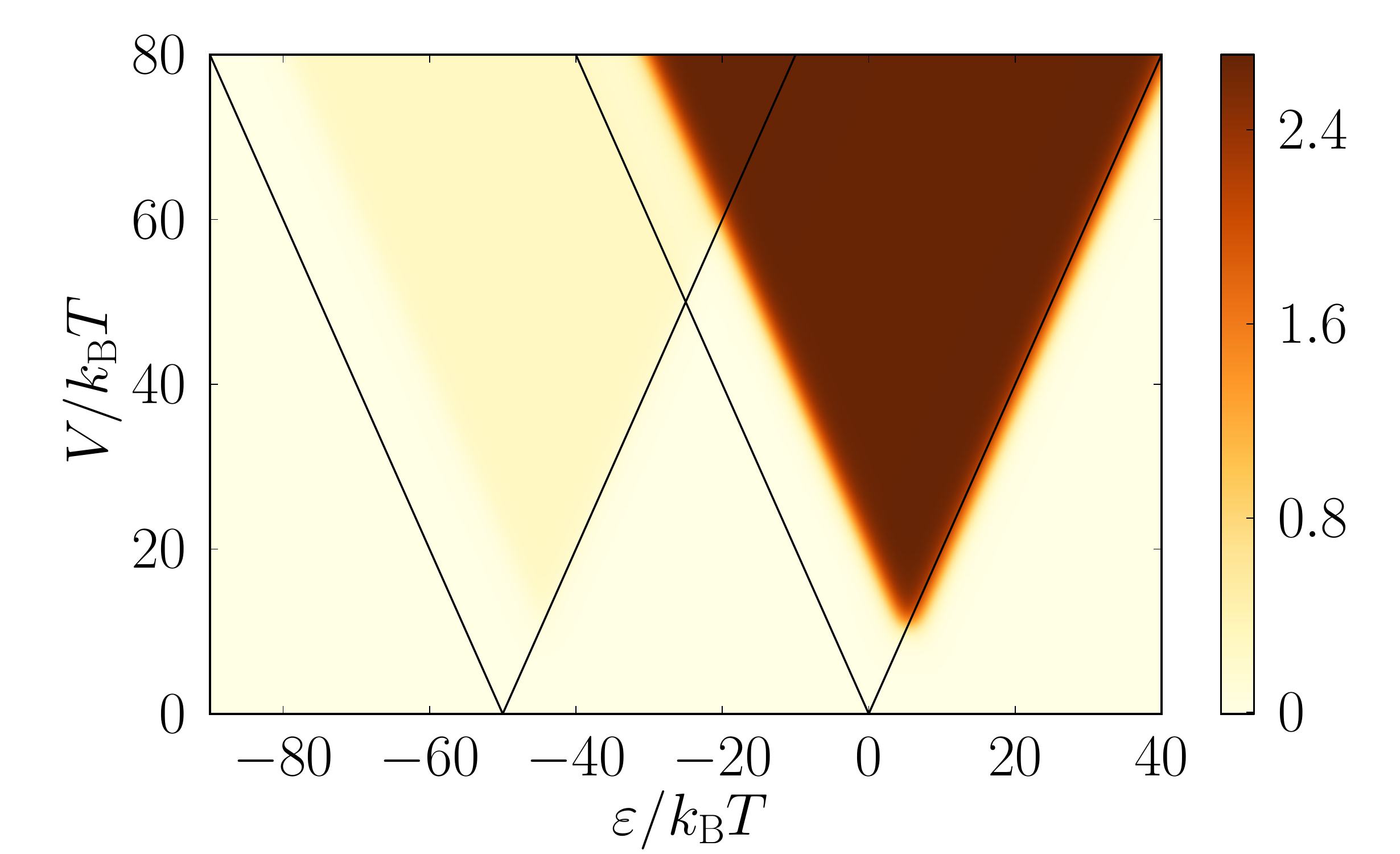}
	\caption{\label{fig:magnonnumber}(Color online) Difference between the average number of magnons in the source and drain lead in the parallel (upper panel) and antiparallel (lower panel) configuration for $\tau=100/\Gamma$, $p=0.9$ and $T_B=T$. Other parameters are the same as in Fig.~\ref{fig:conductancecut}.}
\end{figure}
As the tunneling processes in the quantum-dot spin valve can emit and absorb magnons, they will give rise to a nonequilibrium magnon distribution in the source and drain lead. As we discussed in Sec.~\ref{ssec:processes}, the rate for absorbing a magnon in the source lead is enhanced compared to the rate for emitting a magnon. Hence, the average magnon number in the source will be reduced compared to the equilibrium distribution. For the drain lead, the situation is reversed. Here, the rate for emitting a magnon is enhanced resulting in an increased magnon number compared to equilibrium. In Fig.~\ref{fig:magnonnumber} we plot the difference between the average magnon number in the source and drain lead as a function of gate and bias voltage. We find that indeed the number of magnons in the drain lead is larger than the one in the source lead. For parallel magnetizations, this effect occurs for all gate voltages, while for antiparallel configurations the effect occurs only for transport through the states $\ket{0}$ and $\ket{\sigma}$ as only in this case the magnons can help to overcome the spin blockade on the dot. As they do this very efficiently, the nonequilibrium effects are more pronounced here compared to the parallel configuration giving rise to a larger deviation between the average magnon number in source and drain.

\subsection{\label{ssec:pumpcurr}Magnon-driven electron transport}
\begin{figure}
	\includegraphics[width=.48\textwidth]{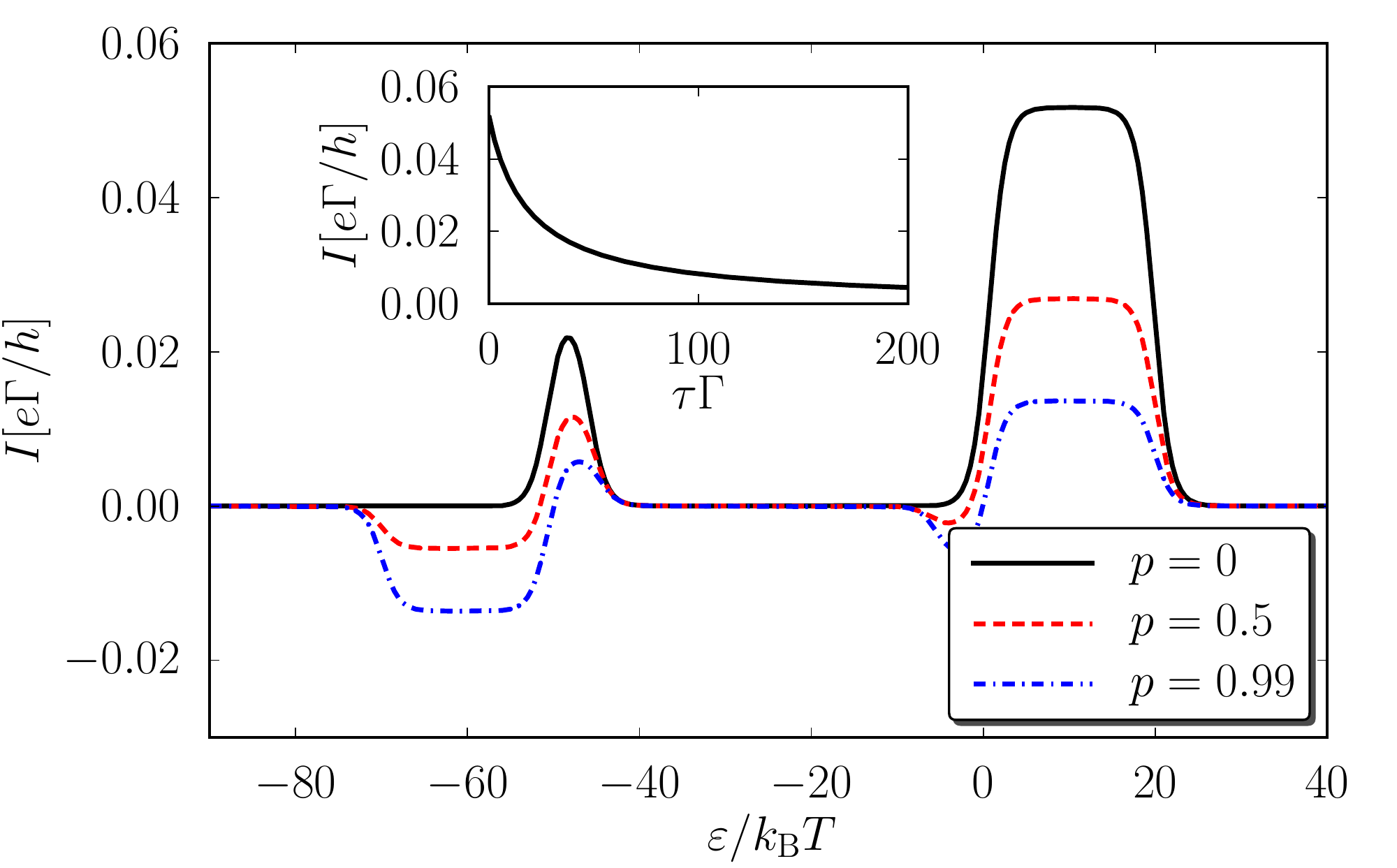}
	\caption{\label{fig:pumpcurr}(Color online) Current at zero bias voltage vs. level position for $U=50\kB T$, $\omega=20\kB T$, $\Gamma_\text{L}=\Gamma_\text{R}$, $\tau=1/\Gamma$, $\lambda=1/3$, $T_B=10T$ at different polarizations. Inset: Increased relaxation time leads to a reduction of the magnon-driven current (shown here the peak value at $\varepsilon=10\kB T$) since it allows a stronger cooling of the magnons.}
\end{figure}
In the following, we show how an asymmetric coupling to the magnons in the left and right electrode can lead to a completely spin-polarized, magnon-driven current  at zero bias voltage. A finite current quite generically requires a breaking of detailed balance, $W_{\xi\xi'}=e^{\beta(E_\xi-E_{\xi'})}W_{\xi'\xi}$. While usually detailed balance is broken by a finite applied bias voltage, here we break detailed balance by different equilibrium temperatures for the magnons and electrons, $T_\text{B}\neq T$. Similar effects occur quite generally when the system couples asymmetrically to external fields and breaks detailed balance.~\cite{bruder_charging_1994,schoeller_transport_1997} Experimentally, this mechanism has been realized by coupling microwaves to the gate electrode defining a quantum dot in a two-dimensional electron gas~\cite{kouwenhoven_quantized_1991,kouwenhoven_observation_1994,kouwenhoven_photon-assisted_1994,oosterkamp_photon-assisted_1996} and in a carbon nanotube~\cite{meyer_photon-assisted_2007} giving rise to photon-assisted tunneling.

For the quantum-dot spin valve, there are different ways to achieve an asymmetric coupling to the magnonic degrees of freedom. We can either choose asymmetric tunnel couplings, $\Gamma_L\neq\Gamma_R$, different couplings to the spin waves $\lambda_r$ or different polarizations $p_r$.
In the following, we choose a system with one ferromagnetic and one normal lead, i.e., we have magnons only in the left lead and $p_\text{R}=0$. In order to violate detailed balance, we choose the equilibrium distribution of the magnons to have a larger temperature than the electrons in the leads.
We note that to describe magnon-assisted tunneling, we do not have to keep track of the magnon numbers in the leads explicitly. Instead, we could also integrate out the magnons and described them as an additional bath with temperature $T_\text{B}$.

In Fig.~\ref{fig:pumpcurr}, we show the magnon-assisted current for vanishing transport voltage as a function of level position for different polarizations. For $0<\varepsilon<\omega_b$, a completely spin-polarized current flows from the ferromagnet into the normal lead. A spin down electron in the ferromagnet can absorb a magnon and tunnel onto the empty dot. Subsequently, the electron can leave the dot either into the ferromagnet or into the normal metal giving rise to a net current into the normal metal. Increasing the polarization enhances the probability of the spin-flip processes that populate the quantum dot. Furthermore, the tunneling rates for the spin-flipped electron back into the ferromagnet are decreased. Hence, with increasing the polarization of the ferromagnet we find an increased magnon-assisted current.

Similarly, for $-U-\omega_b<\varepsilon<-U$, a completely spin-polarized current flows from the normal lead into the ferromagnet. In this case, a spin up electron can leave the quantum dot into the ferromagnet by absorbing a magnon. Afterward, electrons from the left as well as the right lead can tunnel onto the dot, yielding a net current from the normal to the ferromagnetic lead. Since the rate for the absorption processes now is proportional to $1-p$, these processes become strongly suppressed for large polarizations.

In addition to these current plateaus whose width scales with the magnon energy, there are additional peaks at $\varepsilon=0$ and $\varepsilon=-U$ whose width is given by the electron temperature. They occur since at resonance with the leads, the system may flip the spin on the quantum dot which tends to align as to inhibit magnon-absorption processes, thereby making the processes described above possible again and yielding a finite current.

As just illuminated, the key ingredient to the zero-bias current at $T_\text{B}>T$ is the absorption of magnons (similarly, for $T_\text{B}<T$ it would be the emission of magnons). This explains why the height of the current plateaus scales linearly with the average number of magnons. Furthermore, it explains why the current is reduced when the relaxation time is increased (see inset of Fig. \ref{fig:pumpcurr}). In this case, the system absorbs magnons in the processes discussed above. Since it takes more time to relax to the equilibrium magnon distribution now, the average number of magnons is reduced and therefore also the current is reduced.

If we compare our results for the magnon-driven current to the case of a current driven by photon-assisted tunneling, we find that although the basic mechanism of absorbing bosons to gain energy is the same in both cases, there are nevertheless some striking differences. First, in contrast to the photon case, there are no processes which involve the absorption of more than one magnon which is a result of angular momentum conservation during tunneling, cf. the discussion in Sec.~\ref{sec:technique}. Second, while the photons drive an unpolarized current, the magnons yield a \emph{fully} spin-polarized current since in absorption processes they couple only to minority spins. This also leads to a breaking of particle-hole symmetry for finite values of $p_L$.

\subsection{Current noise}
So far we discussed the current and conductance of a quantum-dot spin valve with spin wave degrees of freedom. We now turn to the discussion of the current noise which can provide additional information about the transport processes occurring in the system.

\subsubsection{Zero-frequency noise}
\begin{figure}
	\includegraphics[width=.48\textwidth]{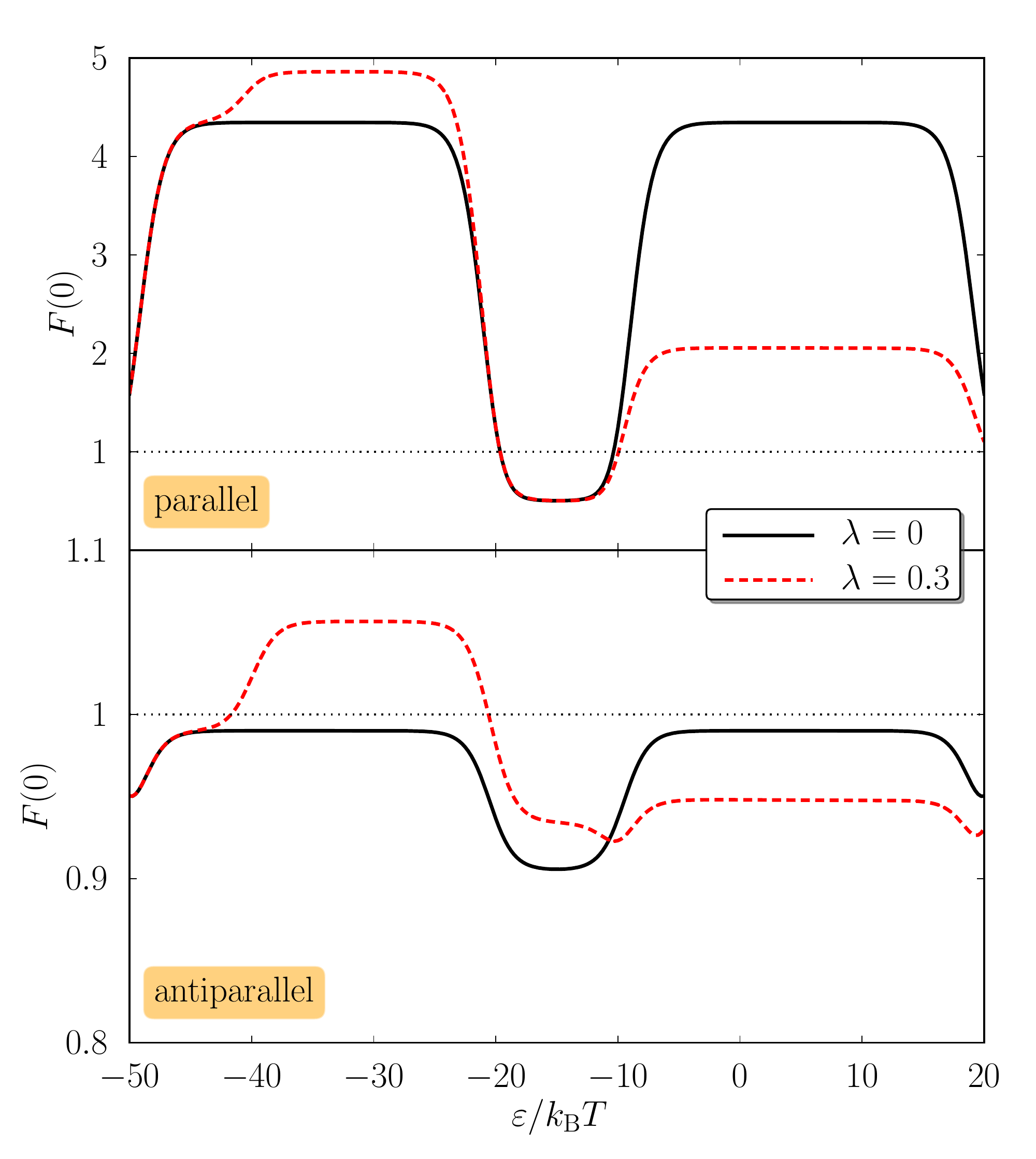}
	\caption{\label{fig:F(0)}(Color online) Fano factor as a function of the level position for parallel (upper panel) and antiparallel (lower panel) magnetizations. In both cases the Fano factor can be increased or decreased by the coupling to the magnons. Parameters are $V=40\kB T$, $U=50\kB T$, $\omega_b=10\kB T$, $\Gamma_\text{L}=\Gamma_\text{R}$, $p=0.9$, $\tau=2/\Gamma$ and $T_B=T$.}
\end{figure}
\begin{figure}
	\includegraphics[width=.48\textwidth]{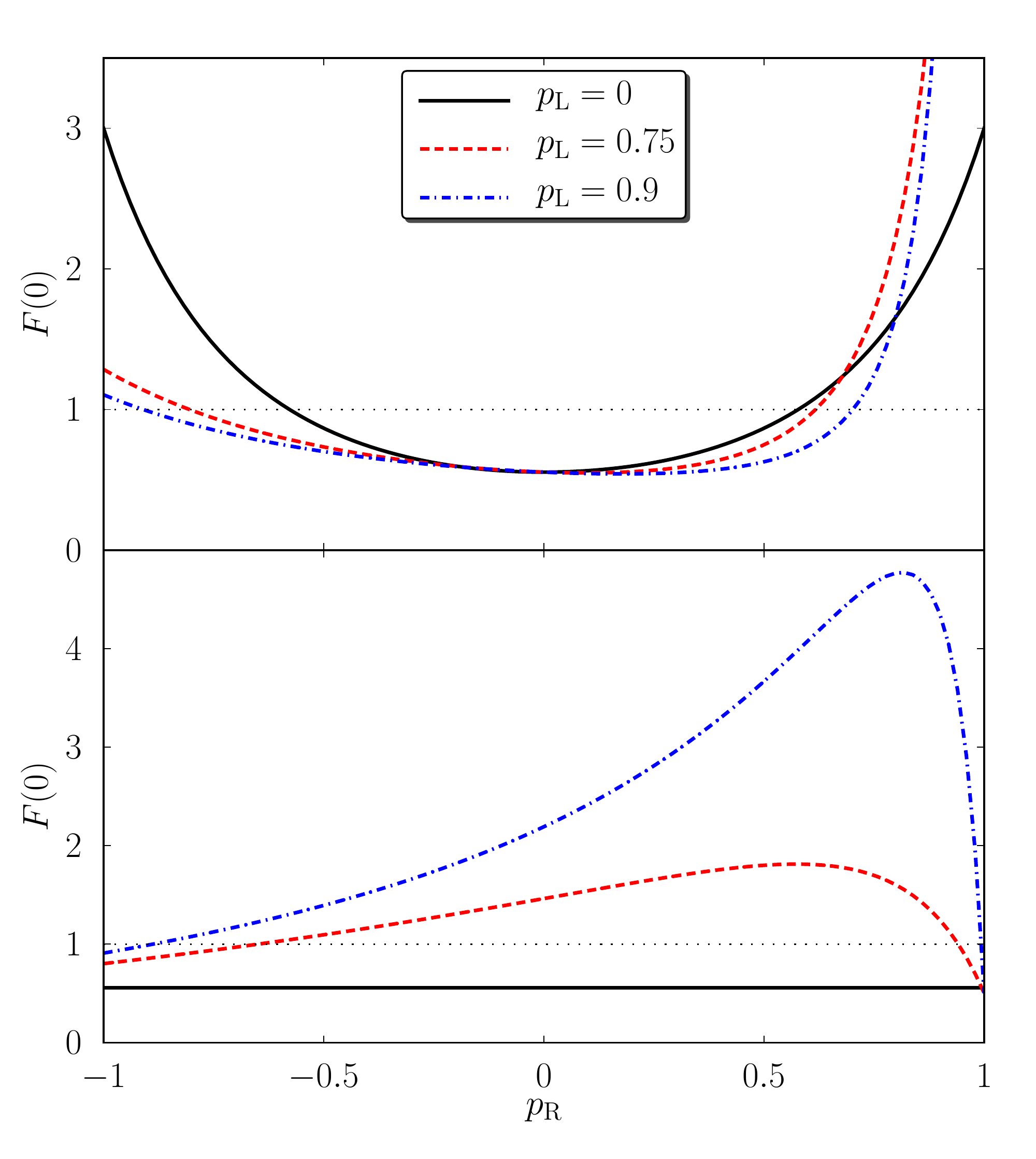}
	\caption{\label{fig:F(0)analytic}(Color online) Zero-frequency Fano factor for a quantum dot symmetrically coupled to collinear magnetized leads as a function of the polarization of the right lead for different values of the polarization of the left lead. In the upper panel, transport takes place through the empty and singly occupied dot only, while in the lower panel, it takes place through the singly and doubly occupied dot only.}
\end{figure}

We start with the discussion of the zero-frequency noise. As shown in Fig.~\ref{fig:F(0)}, where we plot the zero-frequency Fano factor $F(\omega=0)=S(\omega=0)/(2eI)$ as a function of the level position, for sufficiently high polarizations, a quantum dot coupled to ferromagnetic leads with parallel magnetizations will exhibit super-Poissonian Fano factors in the regime where
only two charge states of the dot contribute to transport. This is due to a dynamical spin blockade,~\cite{cottet_dynamical_2004,cottet_positive_2004,cottet_positive_2004-1,belzig_full_2005,elste_transport_2006,buka_shot_1999,buka_current_2000,braun_frequency-dependent_2006,weymann_transport_2007} where minority spin electrons block transport due to their longer dwell time on the dot. Thereby, they chop the current into bunches of majority spin electrons.

If the excitation of spin waves is energetically possible, the Fano factor still remains super-Poissonian.
However, it is significantly reduced when transport takes place through the states $\ket{0}$ and $\ket{\sigma}$ (right plateau in Fig.~\ref{fig:F(0)}) while it is slightly increased for transport through the states $\ket{\sigma}$ and $\ket{d}$ (left plateau in Fig.~\ref{fig:F(0)}). The reduction of the Fano factor on the right plateau can be understood qualitatively by taking into account that the emission of a magnon in the drain lead gives the possibility for a minority spin to leave the dot. Therefore, magnonic processes reduce the waiting time between bunches of majority spin electrons and reduce the Fano factor in consequence. On the left plateau, due to the magnonic processes, there are now two ways to get into the blocking state. Either a spin down electron tunnels to the drain in a normal tunneling event or it does so in a spin-flip process by exciting a spin wave. Hence, the bunching effect becomes stronger in this region and the Fano factor is increased.

In order to gain a more quantitative understanding of the influence of magnons on the Fano factor, we define an effective polarization of the leads which takes into account the magnonic processes.
In the absence of spin waves, the ratio between the rate for transferring a spin-up electron from the dot into the lead and the rate for transferring any electron from the dot to the lead equals $(1+p)/2$. In the presence of spin waves, there are new processes (cf. the discussion in Sec.~\ref{ssec:processes}) which have a different dependence on the polarizations. We therefore define effective polarizations that take into account the presence of the new magnonic processes via
\begin{widetext}
\begin{align}
	\frac{1+p_\text{L,eff}}{2}&=\frac{W_{\up n_\text{L}n_\text{R},0 n_\text{L}n_\text{R}}+W_{\up n_\text{L}+1n_\text{R},0n_\text{L}n_\text{R}}}{W_{\up n_\text{L}n_\text{R},0 n_\text{L}n_\text{R}}+W_{\down n_\text{L}n_\text{R},0n_\text{L}n_\text{R}}+W_{\up n_\text{L}+1n_\text{R},0n_\text{L}n_\text{R}}+W_{\down n_\text{L}-1n_\text{R},0 n_\text{L}n_\text{R}}},\\
	\frac{1+p_\text{R,eff}}{2}&=\frac{W_{0n_\text{L}n_\text{R},\up n_\text{L}n_\text{R}}+W_{0n_\text{L}n_\text{R}-1,\up n_\text{L}n_\text{R}}}{W_{0n_\text{L}n_\text{R},\up n_\text{L}n_\text{R}}+W_{0 n_\text{L}n_\text{R},\down n_\text{L}n_\text{R}}+W_{0 n_\text{L}n_\text{R}-1,\up n_\text{L}n_\text{R}}+W_{0 n_\text{L}n_\text{R}+1,\down n_\text{L}n_\text{R}}}.
\end{align}
\end{widetext}
In the above expressions, we set the magnon numbers that occur in the rates equal to the average magnon number found from the solution of the master equation. This is a reasonable approximation for the parameters chosen here as the system has a very high probability to be in a state with zero magnons.

Hence, the system with spin waves can be interpreted as a quantum dot coupled to ferromagnetic leads with different polarizations $p_L$ and $p_R$ (Note that here different signs of $p_\text{L}$ and $p_\text{R}$ correspond to antiparallel magnetizations and are \emph{not} related to the negative polarizations discussed in Sec.~\ref{sec:model}). For such a system, the Fano factor can be computed analytically and is given by
\begin{equation}\label{eq:F(0)_P}
	F(0)=\frac{(1-2p_\text{L}p_\text{R}+p_\text{R}^2)(5+2p_\text{L}p_\text{R}+p_\text{R}^2)}{(3-2p_\text{L}p_\text{R}-p_\text{R}^2)^2}
\end{equation}
on the right plateau. On the left plateau, the same expression holds with $p_L$ and $p_R$ exchanged. In Fig.~\ref{fig:F(0)analytic}, we show the Fano factor as a function of $p_R$ for various values of $p_L$ in both transport regions. It is interesting to note that for large polarizations, the Fano factor depends quite sensitively on the precise polarization value such that the spin waves can significantly alter the Fano factor. Furthermore, the behavior on the two plateaus is completely different.

Using the picture of effective polarizations, we find that for the parameters of Fig.~\ref{fig:F(0)} where due to $\omega_b\gg T_\text{B}$ and $\tau\sim1/\Gamma$ the average number of magnons is nearly vanishing, we have $p_{\text{L,eff}}=0.90$ and $p_{\text{R,eff}}=0.83$. According to Eq.~\eqref{eq:F(0)_P}, this yields $F(0)=2.1$ for the right plateau in good agreement with the numerically obtained value. Similarly, by exchanging $p_\text{L}$ and $p_\text{R}$ in Eq.~\eqref{eq:F(0)_P}, we find $F(0)=4.8$ for the left plateau, which is also in good agreement with the full calculation.

In the antiparallel case, shown in the lower panel of Fig.~\ref{fig:F(0)}, the zero-frequency Fano factor tends to unity for large polarizations without magnons. In this case, an electron leaving the dot to the drain lead will be easily replaced by another majority electron from the source lead which then blocks transport for a long time. Hence, the tunneling out events become uncorrelated and the Fano factor Poissonian. As in the case of parallel magnetizations, the presence of magnons will result in an effective polarization of the drain smaller than the true polarization. As can be read off from Fig.~\ref{fig:F(0)analytic}, this results in a reduced, sub-Poissonian Fano factor for the right plateau, while it leads to an increased, slightly super-Poissonian Fano factor for the left plateau. For transport at the right plateau, the magnons allow the spin up electron on the quantum dot to leave to the drain lead by flipping its spin. This reduces the waiting times between tunneling events to the drain and therefore makes the Fano factor sub-Poissonian. On the other hand, at the left plateau, after a spin-up electron has left the dot by emitting a magnon in the drain, another spin up electron will enter the dot. In the next step, this scheme will either be repeated or a spin down electron will leave the dot, resulting in the spin-blockaded state. Hence, the magnons can initiate minibunches that lead to slightly super-Poissonian Fano factors.

Since for perpendicular magnetizations, where the Coulomb plateaus become modulated due to the influence of the exchange field,~\cite{braun_frequency-dependent_2006} one does not find any new effects beside the already discussed reduction of the Fano factor for the right and an increase for the left plateau, we do not discuss this case here further.

\subsubsection{Finite-frequency noise}
\begin{figure}
	\includegraphics[width=.48\textwidth]{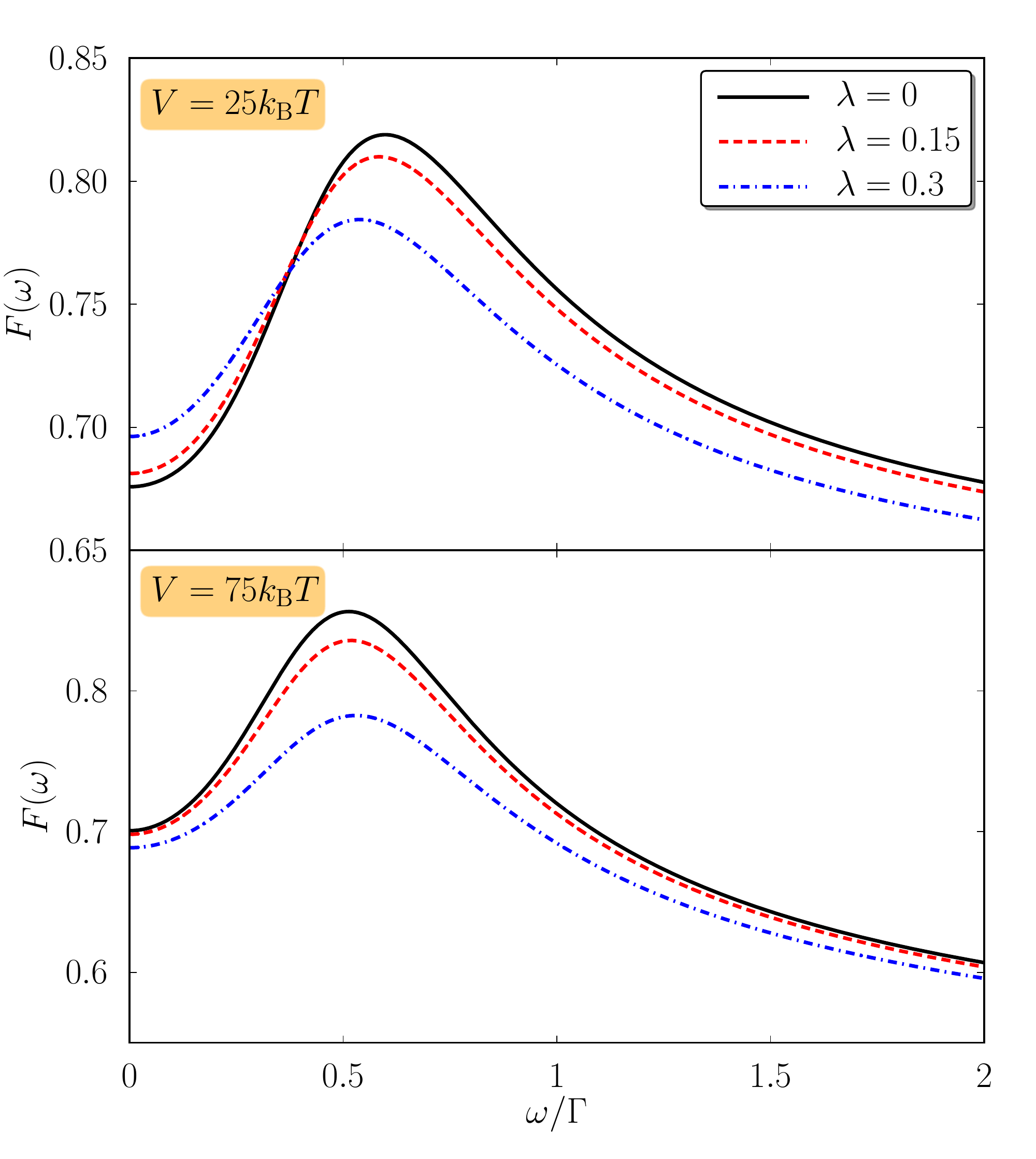}
	\caption{\label{fig:F(omega)}(Color online) Frequency-dependent Fano factor for perpendicular magnetizations for $V=25\kB T$ (upper panel) and $V=75\kB T$ (lower panel). Other parameters are $U=50\kB T$, $\varepsilon=10\kB T$, $\omega_B=10\kB T$, $\Gamma_\text{L}=2\Gamma_\text{R}$, $\tau=1/\Gamma_\text{L}$, $p=0.8$, $T_B=T$.}
\end{figure}
We finally turn to the discussion of the finite-frequency Fano factor $F(\omega)$ for frequencies comparable to the tunnel couplings, $\omega\lesssim\Gamma$. As was discussed in Ref.~\onlinecite{braun_frequency-dependent_2006}, in contrast to the average current which is only sensitive to the average spin on the dot, the finite-frequency current noise provides more direct access to the spin dynamics on the quantum dot. In the following, we discuss how the frequency-dependent Fano factor can be used to detect the magnonic contributions to the exchange field.
The exchange field can only give rise to a precession of an accumulated spin if the exchange field and spin are not collinear to each other. Since the exchange-field contribution of the each lead is parallel to its magnetization direction,
the leads magnetization should not be collinear. To be specific, we choose perpendicularly magnetized leads in the following.

In Fig.~\ref{fig:F(omega)}, we show the frequency-dependent Fano factor for two different bias voltages and 	different coupling strengths to the spin waves. For a vanishing coupling to the magnons, the Fano factor exhibits a peak at the Larmor frequency associated with the exchange field. As the strength of the coupling to the spin waves is increased, the height of the resonance peak is reduced. Furthermore, the position of the resonance peak is shifted as the strength of the exchange field is altered.
Since the magnonic contributions to the exchange field have a different bias dependence as the standard contributions,~\cite{braun_theory_2004} the magnitude and sign of the Larmor frequency shift depends on the bias voltage. Hence, measuring the frequency-dependent Fano factor as a function of bias voltage can provide experimental access to the magnonic exchange field contributions.

\section{\label{sec:conclusions}Conclusions}
We investigated the influence of spin waves excited in the electrodes on transport through a quantum-dot spin valve. We found that the excitation and discexcitation of spin waves gives rise to conductance sidebands whose strength depends on the polarization and the relative orientation of the magnetizations. We, furthermore, found that the transport through the system gives rise to a nonequilibrium occupation of the magnons with an increased magnon number in the drain and a decreased magnon number in the source lead. For a system that couples asymmetrically to the magnons in the source and drain, we showed how magnon-assisted tunneling leads to a completely spin-polarized current without an applied bias voltage. Finally, we studied the current noise. We found that the magnons can increase as well as decrease the Fano factor depending on the dot states involved in transport. Additionally, we showed that the frequency-dependent Fano factor provides access to the magnonic contributions to the exchange field.

\acknowledgments
We acknowledge helpful discussions with Maarten Wegewijs, Sourin Das, Michael Baumgärtel, and Michael Hell. Financial support from DFG via SFB 491 and SPP 1285 and from the European Commission (Grant No. TP7-ICT-2007-C; Project No. 225955 STELE) is gratefully acknowledged.

\appendix
\section{\label{app:diagrams}Diagrammatic rules}
The diagrammatic rules for computing the kernels $W_{\xi_1\xi_1'}^{\xi_2\xi_2'}$ in frequency space are given by
\begin{enumerate}
	\item Draw all topological different diagrams with tunneling lines connecting vertices on either the same or opposite propagators. Assign to the four corners and all propagators states $\ket{\chi, n_\text{L}, n_\text{R}}$ and corresponding energies $E_\chi+\omega_b(n_\text{L}+n_\text{R})$ as well as an energy $\omega$ for every tunneling line.
	\item For each time interval on the real axis confined by two adjacent vertices, assign a resolvent $1/(\Delta E+i\eta)$ where $\Delta E$ is the difference between left and right-going tunneling lines and propagators.
	\item For each tunneling line involving lead $r$, the diagram acquires a factor $\frac{\Gamma_r}{2\pi}f_r^\pm(\omega)$ where the sign on the Fermi function depends on whether the line runs forward ($-$) or backward ($+$) with respect to te Keldysh contour.
	\item For each pair of vertices connected by a tunneling line the diagram is multiplied by $\frac{1+p}{2}\bra{\xi_a'}\tilde c_{r\up}\ket{\xi_a}\bra{\xi_b'}\tilde c_{r\up}^\dagger\ket{\xi_b}+\frac{1-p}{2}\bra{\xi_a'}\tilde c_{r\down}\ket{\xi_a}\bra{\xi_b'}\tilde c_{r\down}^\dagger\ket{\xi_b}$ where $\ket{\xi_a}$ and $\ket{\xi_a'}$ ($\ket{\xi_b}$ and $\ket{\xi_b'}$) are the states that enter and leave the vertex the tunneling line begins (ends) at, respectively. The operators $\tilde c_{r\sigma}^{(\dagger)}$ are defined in Eq.~\eqref{eq:newdotoperators1} and~\eqref{eq:newdotoperators2}. In evaluating the above matrix elements, take into account only terms up to order $\lambda^2$.
	\item Assign a factor of $(-i)(-1)^{a+b}$ where $a$ is the number of vertices on the lower propagator and $b$ is the number of crossings of tunneling lines.
	\item Sum over all leads $r$.
	\item Integrate over all energies of tunneling lines. In the sequential-tunneling regime only one tunneling line is involved. In this case, the frequency integration reduces to a simple application of Cauchy's formula.
	\item For the computation of $\mathbf W^I$ and $\mathbf W^{II}$, one, respectively two tunnel vertices are replaced by current vertices. These give rise to a factor $+1/2$ if they are on the upper (lower) branch of the Keldysh contour and describe the tunneling of an electron into the right (left) or out off the left (right) lead. Otherwise they result in a factor $-1/2$.
\end{enumerate}

\section{\label{app:mastereq}Master equation}
In this appendix, we give expressions for the various functions that enter the master equations \eqref{eq:mastercharge} and \eqref{eq:masterspin}.

Introducing the abbreviations
\begin{widetext}
\begin{eqnarray*}
	r_r&=&(1+p_r)n_rf_r^+(\varepsilon-\omega_b)+(1-p_r)(1+n_r)f_r^+(\varepsilon+\omega_b),\\
	s_r&=&(1+p_r)(1+n_r)f_r^-(\varepsilon+U-\omega_b)-(1-p_r)n_rf_r^-(\varepsilon+\omega_b),\\
	x_r&=&f_r^-(\varepsilon)+f_r^+(\varepsilon+U),\\
	y_r&=&f_r^-(\varepsilon)-f_r^+(\varepsilon+U),\\
	z_{r\pm}&=&(1+p_r)(1+n_r)f_r^-(\varepsilon-\omega_b)\pm(1-p_r)n_rf_r^-(\varepsilon+\omega_b)\\
&&\pm(1+p_r)n_rf_r^+(\varepsilon+U-\omega_b)+(1-p_r)(1+n_r)f_r^+(\varepsilon+U+\omega_b),
\end{eqnarray*}
we can write the quantities $M^{(r)}_{\chi\vec n,\chi'\vec m}$ as matrices in the basis $\ket{0}$, $\ket{1}$, $\ket{d}$:
\begin{eqnarray}
	M^{(r)}_{\vec n,\vec n}&=&\Gamma_r
	\left(\begin{array}{ccc}
		-2(1-\lambda^2 n_r)f_r^+(\varepsilon)-\lambda^2r_r	& (1-\lambda^2 n_r)f_r^-(\varepsilon)		& 0	\\
		2(1-\lambda^2 n_r)f_r^+(\varepsilon)		& -(1-\lambda^2 n_r)y_r+\frac{\lambda^2}{2}z_{r+}	& 2(1-\lambda^2 n_r)f_r^-(\varepsilon+U) \\
		0				& (1-\lambda^2 n_r)f_r^+(\varepsilon+U)		& -2(1-\lambda^2 n_r)f_r^-(\varepsilon+U)-\lambda^2s_r
	\end{array}\right),\\
	M^{(r)}_{\vec n,\vec n-1}&=&\frac{\lambda^2}{2}n_r
	\left(\begin{array}{ccc}
		0 & \Gamma_{r+}f_r^-(\varepsilon-\omega_b) & 0 \\
		2\Gamma_{r-}f_r^+(\varepsilon+\omega_b)  & 0 & 2\Gamma_{r+}f_r^-(\varepsilon+U-\omega_b) \\
		0 & \Gamma_{r-}f_r^+(\varepsilon+U+\omega_b) & 0
	\end{array}\right),\\
	M^{(r)}_{\vec n,\vec n+1}&=&\frac{\lambda^2}{2}(1+n_r)
	\left(\begin{array}{ccc}
		0 & \Gamma_{r-}f_r^-(\varepsilon+\omega_b) & 0 \\
		2\Gamma_{r+}f_r^+(\varepsilon-\omega_b)  & 0 & 2\Gamma_{r-}f_r^-(\varepsilon+U+\omega_b) \\
		0 & \Gamma_{r+}f_r^+(\varepsilon+U-\omega_b) & 0
	\end{array}\right).
\end{eqnarray}
\end{widetext}
The vectors $V_{\chi\vec n}^{(r)}$ and $F_{\chi\vec n\vec m}^{(r)}$ that enter the master equation for the occupations, Eq.~\eqref{eq:mastercharge}, and spin, Eq.~\eqref{eq:masterspin}, respectively, can be written as vectors in the basis $\ket{0}$, $\ket{1}$ and $\ket{d}$ as
\begin{eqnarray}
	V^{(r)}_{\vec n}&=&2p_r\Gamma_r(1-\lambda^2n_r)
	\left(\begin{array}{c}
		f_r^-(\varepsilon)\\
		-y_r+\lambda^2z_{r-}\\
		f_r^-(\varepsilon+U)
	\end{array}\right),\\
	V^{(r)}_{\vec n+1}&=&\lambda^2(1+n_r)
	\left(\begin{array}{c}
		\Gamma_{r-}f_r^-(\varepsilon+\omega_b)\\
		0\\
		\Gamma_{r+}f_r^-(\varepsilon+U-\omega_b)
	\end{array}\right),\\
	V^{(r)}_{\vec n-1}&=&-\lambda^2n_r
	\left(\begin{array}{c}
		\Gamma_{r+}f_r^-(\varepsilon-\omega_b)\\
		0\\
		\Gamma_{r-}f_r^-(\varepsilon+U+\omega_b)
	\end{array}\right).
\end{eqnarray}
and
\begin{eqnarray}
	F^{(r)}_{\vec n\vec n}&=&p_r\Gamma_r(1-\lambda^2n_r)\left(\begin{array}{c}
	f_r^+(\varepsilon) \\
	-\frac{2y_r+z_{r-}}{2} \\
	 -f_r^-(\varepsilon+U)
	\end{array}\right),\\
	F^{(r)}_{\vec n\vec n+1}&=&\frac{\lambda^2}{2}(1+n_r)\left(\begin{array}{c}
	\Gamma_{r+}f_r^+(\varepsilon-\omega_b) \\
	0 \\
	\Gamma_{r-}f_r^-(\varepsilon+U+\omega_b)
	\end{array}\right),\\
	F^{(r)}_{\vec n\vec n-1}&=&\frac{\lambda^2}{2}n_r\left(\begin{array}{c}
	\Gamma_{r-}f_r^+(\varepsilon+\omega_b) \\
	0 \\
	\Gamma_{r+}f_r^-(\varepsilon+U-\omega_b)
	\end{array}\right).
\end{eqnarray}
Finally, we have
\begin{equation}
	G^{(r)}=(1-\lambda^2n_r)\Gamma_rx_r+\frac{\lambda^2}{2}\Gamma_rz_{r+}.
\end{equation}

%\bibliography{/home/bjoern/LaTeX/Bibtex/Meine_Bibliothek.bib}

%Merlin.mbs v4.21 2009-07-09.
%

\end{document}